\documentclass[11pt,twoside,fleqn]{amsart}

\usepackage{etex}
\reserveinserts{50}
\usepackage[english]{babel}
\usepackage{scrextend}
\usepackage{import}

\usepackage[latin1]{inputenc}
\usepackage[T1]{fontenc}
\usepackage{nicefrac}

\usepackage{newpxtext}
\usepackage[scale=0.985,semibold]{cabin}
\usepackage[varg,bigdelims,cmbraces,slantedGreek]{newpxmath}
\usepackage[scaled=1.125,proportional,lightcondensed]{zlmtt}

\usepackage[sans]{dsfont}
\usepackage{yfonts}
\usepackage{nicefrac}
\usepackage{ragged2e}
\usepackage{bm}

\usepackage[babel,german=guillemets,english=american]{csquotes}

\usepackage[
a4paper,
driver=pdftex,
includehead=true,
includefoot=true,
top=1.5cm,
bottom=1.5cm,
inner=3cm,
outer=3cm,
marginparsep=0cm,
marginparwidth=0cm,
footskip=2.75\baselineskip,
headheight=1\baselineskip,
headsep=1.5\baselineskip,
heightrounded,
centering
]{geometry}

\usepackage[
activate={true},
final,
auto=true,
protrusion=true,
expansion=true,
kerning=true,
tracking=true,
spacing=true,
verbose=silent,
babel=true]
{microtype}
\SetTracking[spacing = {25*,166, } ]{ encoding = *, shape = sc }{-5}

\usepackage{setspace}
\allowdisplaybreaks[4]

\usepackage{marginfix} 
\usepackage[multiple,stable,side]{footmisc}

\deffootnote[1.75em]{0em}{0em}{\makebox[1.5em][l]{\textcolor{OliveGreen}{\rule[-0.5mm]{0.75pt}{8.25pt}}\,{\thefootnotemark}}\tiny{\,}}
\deffootnotemark{\tiny{\,}\textcolor{OliveGreen}{\rule[-0.35mm]{0.75pt}{9.75pt}}\raisebox{2.25pt}{\footnotesize\,\thefootnotemark}}

\usepackage[compact,calcwidth,rubberchapters]{titlesec}
\usepackage{titletoc}

\renewcommand\thesection{\arabic{section}}
\renewcommand\thesubsection{\Alph{subsection}}
\renewcommand\theparagraph{\alph{paragraph}}

\titleformat{\section}[hang]
{\Large\color{DFG-blau}\sf\bfseries}
{\makebox[1.5cm][r]{\textcolor{darkgray}{\raisebox{-2.5pt}{\fontsize{24}{24}\selectfont{\thesection.}}}}}
{7pt}
{\Large\sf\color{DFG-blau}\bfseries}

\titleformat{\subsection}[hang]
{\large\sf}
{\makebox[1.5cm][r]{\tikz[baseline=-4.95pt]{
\node[inner 
sep=0pt]{\includegraphics[height=18pt]{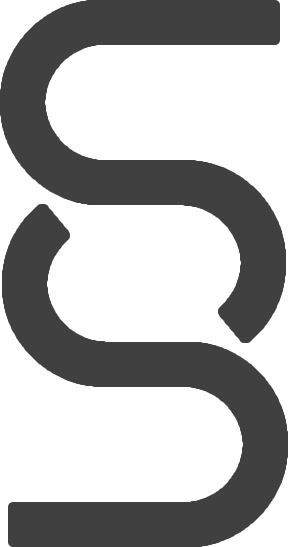}};}\hspace*{1pt}\textcolor{darkgray}{\raisebox{-1.15pt}{\fontsize{18}{18}\selectfont\thesubsection}}}}
{7pt}
{\setstretch{1.0}\color{DFG-blau}}

\titleformat{\paragraph}[runin]
{\normalsize\sf\itshape}
{{\textcolor{darkgray}{(\theparagraph)}}}
{5pt}
{\color{DFG-blau}}

\titlespacing*{\section}
{0pt}
{1\baselineskip plus 0.125\baselineskip minus 0.0625\baselineskip}
{0.\baselineskip plus 0.0625\baselineskip}

\titlespacing*{\subsection}
{0pt}
{0.75\baselineskip plus 0.125\baselineskip minus 0.0625\baselineskip}
{0\baselineskip plus 0.0625\baselineskip}

\titlespacing*{\paragraph}
{0pt}
{0pt}
{2\wordsep}

\usepackage{fancyhdr}
\usepackage{calc}
\usepackage{lastpage}

\fancypagestyle{plain}{\fancyhf{}
\fancyfoot[CO,CE]{\begin{tikzpicture}[overlay]
\draw[line width=1.25pt,-,line cap=round,color=DFG-blau] (-0.375cm-0.5\textwidth,1.075\baselineskip) -- (0.5\textwidth+0.375cm,1.075\baselineskip);
\end{tikzpicture}}
\fancyfoot[RO,LE]{\textcolor{darkgray}{\textcolor{darkgray}{\small\sf Page \thepage\@ of \pageref{LastPage}}}}
\fancyfoot[RE,LO]{\textcolor{darkgray}{\small\sf \today}}}

\renewcommand{\sectionmark}[1]{\markleft{#1}}

\usepackage{etoolbox}
\makeatletter
\patchcmd{\ttl@select}{\strut}{}{}{}
\patchcmd{\ttlh@hang}{\strut}{}{}{}
\patchcmd{\ttlh@hang}{\strut}{}{}{}
\makeatother

\makeatletter
\patchcmd{\ttl@straight@ii}{\vspace{\@tempskipb}}{\vskip \@tempskipb}{}{}
\makeatother

\usepackage[maxfloats=50]{morefloats}
\usepackage{graphicx}
\usepackage{multirow,longtable,booktabs,tabularx,tabulary}
\usepackage[usenames,dvipsnames,table]{xcolor}

\usepackage{tikz}

\usepackage{pifont}
\usepackage[misc]{ifsym}
\usepackage{bbding}

\usepackage{marvosym}
\usepackage{dictsym}
\usepackage[htt]{hyphenat}

\usepackage{mathtools}
\usepackage{nccmath}
\usepackage{enumitem}
\usepackage{soul}

\usepackage{multicol}
\usepackage[noorphans,vskip=-0.25\baselineskip plus 0.125\baselineskip minus 0.0625\baselineskip]{quoting}

\usepackage{fontawesome}
\pdfmapfile{=fontawesome.map}

\usepackage{caption}
\usepackage{floatrow}

\addto{\captionsngerman}{%

}

\addto{\captionsenglish}{%

}

\usepackage[
backend=bibtex,
natbib=true,
autocite=superscript,
style=numeric-comp,
ibidtracker=true,
idemtracker=true,
maxnames=99,
doi=true,
url=true,
isbn=false,
sorting=none,
autopunct=false,
backref,
mcite]
{biblatex}

\DeclareFieldFormat{pages}{#1}
\DeclareFieldFormat{authortype}{\textsc{#1}}
\DeclareFieldFormat{author}{\textsc{#1}}
\DeclareFieldFormat{title}{\mkbibquote{\normalfont #1}}
\DeclareFieldFormat{journaltitle}{\normalfont\itshape #1}
\DeclareFieldFormat[article]{volume}{\mkbibbold{#1}} 
\DeclareFieldFormat{booktitle}{\normalfont #1}
\DeclareFieldFormat{year}{\normalfont\useosf{#1}}
\DeclareFieldFormat{note}{\normalfont\itshape #1}

\renewbibmacro{in:}{\ifentrytype{article}{}{\printtext{\bibstring{in}\intitlepunct}}}

\DeclareFieldFormat{postnote}{\mkpageprefix[pagination]{#1}}

\AtBeginDocument{%

}

\DefineBibliographyStrings{german}{%
  references = {Zitierte Literatur},
}

\DeclareFieldFormat{labelnumberwidth}{\makebox[1.95em][r]{\makebox[1.75em][c]{\color{DFG-blau}\sf\small{#1}}\;\color{DFG-blau}\rule[-1.25mm]{1pt}{13.75pt}}
\hspace*{-8pt}}

\DeclareCiteCommand{\supercite}
[\mkbibsuperscript]%
{\usebibmacro{cite:init}%
\iffieldundef{prenote}%
{}%
{\BibliographyWarning{Ignoring prenote argument}}%
\iffieldundef{postnote}%
{}%
{\BibliographyWarning{Ignoring postnote argument}}%
{\hspace*{0.5pt}{\raisebox{-0.15pt}{\textcolor{DFG-blau}{\scriptsize\faChevronCircleRight}}}\hspace*{-0.75pt}}
}%
{\sf\scriptsize\bfseries\usebibmacro{citeindex}%
\usebibmacro{cite:comp}}
{}%
{\usebibmacro{cite:dump}
\hspace*{-3.5pt}
}%

\newbibmacro{string+doiurlisbn}[1]{%
  \iffieldundef{doi}{%
    \iffieldundef{url}{%
      \iffieldundef{isbn}{%
        \iffieldundef{issn}{%
          #1%
        }{%
          \href{http://books.google.com/books?vid=ISSN\thefield{issn}}{#1}%
        }%
      }{%
        \href{http://books.google.com/books?vid=ISBN\thefield{isbn}}{#1}%
      }%
    }{%
      \href{\thefield{url}}{#1}%
    }%
  }{%
    \href{http://dx.doi.org/\thefield{doi}}{#1}%
  }%
}

\newcommand*\extlinktext{\tikz[baseline=-3pt]{\node[inner sep=0pt]{\includegraphics[height=8pt]{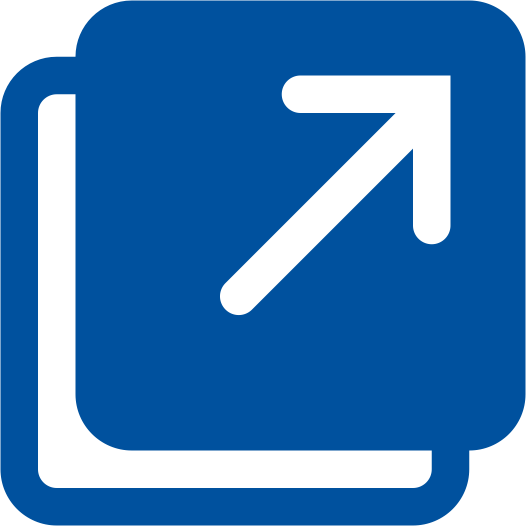}};}}

\DeclareFieldFormat*{url}{\usebibmacro{string+doiurlisbn}\extlinktext\nolinebreak[4]\nopunct}
\DeclareFieldFormat*{doi}{\usebibmacro{string+doiurlisbn}\extlinktext\nolinebreak[4]\nopunct}
\DeclareFieldFormat*{urldate}{}

\DeclareCiteCommand{\citenum}
  {}
  {\printfield{labelnumber}}
  {}
  {}

\usepackage[
colorlinks,
hyperfigures,
breaklinks=true,
raiselinks,
nesting,
hyperindex,
pdfview=Fit,
bookmarksnumbered,
bookmarksopen,
hyperfootnotes=false,
pdfstartview=FitH,
linkcolor=DFG-blau,
citecolor=DFG-blau,
urlcolor=fu-rot,
filecolor=yellow,
pdfauthor={Thomas Grohmann},
pdfcreator={Thomas Grohmann},
pdftitle={On the impossibility of frozen nuclei},
]{hyperref}

\renewcaptionname{english}\figureautorefname{Fig.}
\renewcaptionname{english}\equationautorefname{Eq.}
\renewcaptionname{english}\sectionautorefname{Section}
\renewcaptionname{english}\subsectionautorefname{Subsection}

\usepackage{catoptions}
\makeatletter
\def\figureautorefname{figure}

\def\Autoref#1{%
  \begingroup
  \edef\reserved@a{\cpttrimspaces{#1}}%
  \ifcsndefTF{r@#1}{%
    \xaftercsname{\expandafter\testreftype\@fourthoffive}
      {r@\reserved@a}.\\{#1}%
  }{%
    \ref{#1}%
  }%
  \endgroup
}
\def\testreftype#1.#2\\#3{%
  \ifcsndefTF{#1autorefname}{%
    \def\reserved@a##1##2\@nil{%
      \uppercase{\def\ref@name{##1}}%
      \csn@edef{#1autorefname}{\ref@name##2}%
      \autoref{#3}%
    }%
    \reserved@a#1\@nil
  }{%
    \autoref{#3}%
  }%
}
\makeatother
\def\be{\begin{equation}}
\def\ee{\end{equation}}
\def\bse{\begin{subequations}}
\def\ese{\end{subequations}}
\def\beo{\begin{equation*}}
\def\eeo{\end{equation*}}
\def\bea{\begin{eqnarray}}
\def\eea{\end{eqnarray}}
\def\beao{\begin{eqnarray*}}
\def\eeao{\end{eqnarray*}}
\def\beqo{\begin{quote}}
\def\enqo{\end{quote}}
\def\ben{\begin{enumerate}}
\def\een{\end{enumerate}}
\def\bit{\begin{itemize}}
\def\eit{\end{itemize}}
\def\bed{\begin{description}}
\def\eed{\end{description}}
\def\berale{\begin{raggedright}}
\def\eerale{\end{raggedright}}

\def\tssc#1{\textsuperscript{#1}}

\def\mal#1{\mathcal #1}
\def\mak#1{\mathfrak #1}

\definecolor{peru}{RGB}{205,133,63} 
\definecolor{meergruen}{RGB}{46,139,87}
\definecolor{dunkel-blau}{RGB}{0,0,139}
\definecolor{dunkel-magenta}{RGB}{139,0,139}
\definecolor{rot}{rgb}{0.5,0,0}
\definecolor{grun}{rgb}{0,0.3,0}
\definecolor{pale-brown}{RGB}{152,118,54}
\definecolor{copper}{RGB}{184,115,51}
\definecolor{bronze}{RGB}{205,127,50}
\definecolor{wheat}{RGB}{245,222,179}
\definecolor{tan}{RGB}{210,180,140}
\definecolor{orche}{RGB}{204,119,34}
\definecolor{corn}{RGB}{251,236,93}
\definecolor{golden-yellow}{RGB}{255,223,0}
\definecolor{dark-olive}{RGB}{85,107,47}
\definecolor{dark-orange}{RGB}{238,173,14}
\definecolor{dark-brown}{RGB}{92,64,51}
\definecolor{nwpurple}{RGB}{82,0,99}
\definecolor{fu-blau}{RGB}{0,51,102}
\definecolor{fu-gruen}{RGB}{153,204,0}
\definecolor{fu-gruen-2}{RGB}{122,173,20}
\definecolor{fu-rot}{RGB}{204,0,0}
\definecolor{fu-orange}{RGB}{255,153,0}
\definecolor{DFG-blau}{RGB}{0,81,158}
\definecolor{DFG-rot}{RGB}{229,53,23}
\definecolor{DFG-gruen}{RGB}{122 ,181,29}
\definecolor{DFG-gelb}{RGB}{250,186,0}

\newcolumntype{C}[1]{>{\centering\arraybackslash}m{#1}}
\newcolumntype{R}[1]{>{\raggedleft\arraybackslash}m{#1}}
\newcolumntype{L}[1]{>{\raggedright\arraybackslash}m{#1}}
\newcolumntype{Z}[1]{>{\arraybackslash}p{#1}}

\newtagform{brackets2}[\color{DFG-blau}\sf]{\textcolor{DFG-blau}{\sf Eq.{\tiny\,}}{}\sf}{{}}
\usetagform{brackets2}

\newtagform{brackets3}[\color{DFG-blau}\sf]{\hspace*{18pt}\textcolor{DFG-blau}{\sf Eq.{\tiny\,}}{}\sf}{{}}

\newcommand*\TightFrame[1]{{\tikz[baseline=-3.95pt]{\node[rounded corners=1.75pt,draw,color=darkgray, inner sep=0pt,thick=0pt,minimum height=13pt,minimum 
width=13pt,fill=White,text centered,line width=1pt,fill=DFG-blau!85!,text=White] {#1};}}}

\SetLabelAlign{Center}{\hfil#1\hfil}

\newcommand\NamEs{Grohmann}
\newcommand\TitLe{On the impossibility of frozen nuclei}

\begin{document}
\setlength{\topskip}{1\baselineskip plus 0.125\baselineskip minus 0.0625\baselineskip}
\setlength{\parskip}{0.75\baselineskip plus 0.05\baselineskip minus 0.05\baselineskip}
\setlength{\parindent}{0pt}
\setlength{\footnotesep}{0.75\baselineskip plus 0.125\baselineskip minus 0.075\baselineskip}
\flushbottom

\setlength\mathindent{4.25em}

\selectlanguage{english} 
\pagestyle{fancy}

\thispagestyle{plain}
\renewcommand{\sectionmark}[1]{\markleft{#1}}

\vspace*{0.75\baselineskip}

\setstretch{2.95}
{\Huge\sf\bfseries\fontsize{32}{32}\selectfont\textcolor{DFG-blau}{On the impossibility\\ of frozen nuclei}}\\[-0.25\baselineskip]

\setstretch{1.125}
{\large\sf\scshape Thomas Grohmann}

\onehalfspacing
\bit
[leftmargin=20pt,
rightmargin=0pt,
labelsep=5pt,
topsep=0\baselineskip,
font=\small\bf,
itemsep=0.075\baselineskip plus 0.0625\baselineskip minus 0.0625\baselineskip]
\item[\TightFrame{\hspace*{0.5pt}\scriptsize\faUniversity}] 
{\sf Institut f\"ur Chemie und Biochemie, Freie Universität Berlin}

\item[\TightFrame{\scriptsize\faEnvelope}] 
{\sf thomas.grohmann\faAt fu-berlin.de}

\item[\TightFrame{\hspace*{0.35pt}\small\faHistory}]
{\sf Second, edited and slightly modified version;} {\sf\today}
\eit

\tikz[overlay]{\node[anchor=east,inner sep=0pt] at (\textwidth,1.5) {\includegraphics[width=3cm]{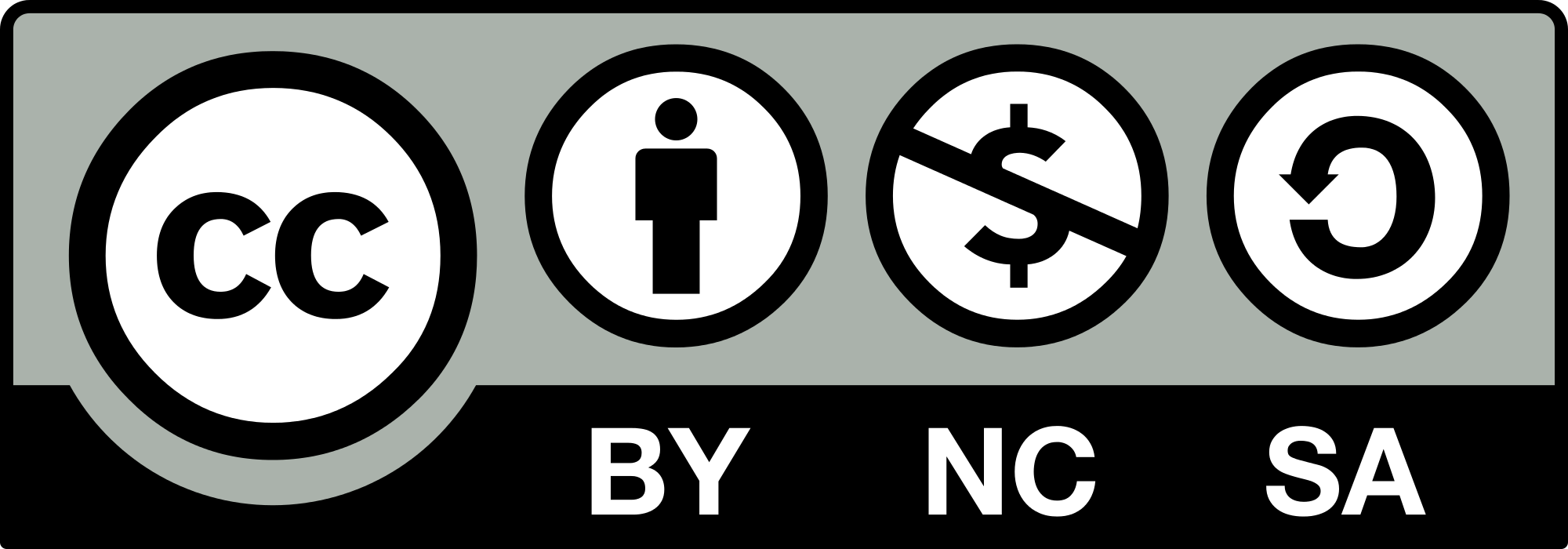}};}

\noindent
\vspace{-3.75\baselineskip plus 0.125\baselineskip}
\singlespacing

\begin{center}
{\tikz[overlay,centered]{\draw [line width=1pt,-,line cap=round,color=darkgray] (-0.5\textwidth-0.125cm,0) -- (0.5\textwidth+0.125cm,0);}}
\end{center}
\vspace*{-0.25\baselineskip}

{\small Many molecular \textquote{quantum} theories, like \textquote{quantum chemistry}, conceal that they are actually quantum-classical approaches---they treat one set of 
molecular degrees of freedom classically while the remaining degrees of freedom follow the laws of quantum mechanics. We show that the prominent 
\textquote{frozen-nuclei approximation}, which is often used in molecular control communities, is a further example for such theory reduction: It treats the nuclei of the molecule 
as classical particles. Here, we demonstrate that the ignorance about the quantum nature of nuclei has far-reaching consequences for the theoretical description of molecules. We 
analyse the symmetry of oriented and aligned rigid molecules with feasible permutations of identical nuclei and show: The presumption of fixed nuclei corresponds to a localized 
state that is impossible to create if the existence of stable nuclear spin isomers is a justifiable assumption for the controlled molecule. The results of studies on molecules 
containing identical nuclei have to be re-evaluated and properly anti-symmetrised, because for such molecules the premise of frozen nuclei is inherently wrong: Molecular wave 
functions have to obey the spin-statistics theorem twice.}

\noindent
\vspace{-2.25\baselineskip plus 0.125\baselineskip}

\begin{center}
{\tikz[overlay,centered]{\draw [line width=1pt,-,line cap=round,color=darkgray] (-0.5\textwidth-0.125cm,0) -- (0.5\textwidth+0.125cm,0);}}
\end{center}

\setstretch{1.15}
\section{Frozen nuclei in molecular control}
\label{sec:frozen-nuc}
It seems as if the laser control of molecular processes is a story of success. Steering molecular motions with laser pulses on pico-, femto-, and attosecond time-scales does not 
only allow for many interesting applications, such as the control of molecular orientation and alignment,\autocite{Friedrich.1991,Stapelfeldt.2003,Seideman.2005,Lemeshko.2013} 
the control of charge-transfer,\autocite{Vrakking.2014,Ramasesha.2016} or the design of molecular switches and molecular 
rotors.\autocite{Kottas.2005,Feringa.2017,Stoddart.2017} It also puts some long-standing debates in the theory of chemistry back into spotlight: Can we measure orbitals and 
what, if anything, do they actually 
mean?\autocite{Ogilvie.1990,Scerri.2000,Scerri.2001,Zuo.2001,Schwarz.2006,Ostrovsky.2005,Labarca.2010,Mulder.2010,Mulder.2011,Ogilvie.2011,Autschbach.2012,Villani.2017}
Do molecules have a structure, and if so, can we measure 
it?\autocite{Woolley.1976,Primas.1983,Weininger.1984,Woolley.1985,Primas.1985,Sutcliffe.1992,Ramsey.1997,Primas.1998,Hendry.2010,Matyus.2011,Sutcliffe.2012,Ochiai.2017,Ghibaudi.2019}
How do electrons move during chemical reactions?\autocite{Kling.2008,Bredtmann.2015} And what follows from all of this for our understanding of chemistry? Addressing such 
questions is not only relevant for developing theories of chemistry and physics to a sophisticated level. It also helps us to better understand the complicated relationship of 
chemistry and quantum theory.\autocite{Primas.1983,Scerri.2000,Weisberg.2016,Hettema.2017}

Attosecond scientists in particular have focussed on giving new answers to these fundamental questions. In numerous studies, they claim to have measured what standard 
interpretations of quantum theory conclude is impossible to observe: Orbitals and molecular structures can be seen in time-resolved experiments for many different systems, 
attosecond scientists have repeatedly 
reported.\autocite{Itatani.2004,Bucksbaum.2007,Kling.2008,Haessler.2010,Salieres.2012,Spanner.2013,Diveki.2013,Vrakking.2014,Ramasesha.2016,Peng.2019} Critical 
comments on such measurements are 
rare,\autocite{Ogilvie.1990,Scerri.2000a,Scerri.2001,Ostrovsky.2005,Schwarz.2006,Labarca.2010,Mulder.2010,Mulder.2011,Ogilvie.2011,Autschbach.2012}
and they usually address only the problem that any exact electronic wave function cannot be decomposed into (anti-symmetrised) products of one-electron wave functions: 
Because the \textquote{orbital approximation} is incorrect at any level of theory, there is nothing to 
observe in accurate experiments.\autocite{Ogilvie.1990,Scerri.2000,Scerri.2001,Labarca.2010,Ogilvie.2011} Moreover, wave functions cannot be measured directly, the popular
Copenhagen interpretation of quantum theory dictates to us.\autocite{Schlosshauer.2005} How, then, can we image one-electron wave functions at 
all?\autocite{Schwarz.2006,Schwarz.2001}

In this paper, we lengthen the list of arguments why some results of attosecond research should lead sceptics to reasonable doubt. We focus, however, on a different aspect: the 
classical treatment of the nuclei. Here, we argue that the \textquote{frozen-nuclei approximation}, which is widely employed by 
atto-scientists,\autocite{Itatani.2004,Bucksbaum.2007,Kling.2008,Haessler.2010,Salieres.2012,Spanner.2013,Diveki.2013,Vrakking.2014,Ramasesha.2016,Peng.2019} is physically 
incorrect for molecules with identical nuclei. Because it is an inherently quantum-classical theory, treating the nuclei as classical objects by fixing them in 
space,\autocite{Primas.1983,Primas.1998} the {frozen-nuclei approximation} not only violates the indistinguishability of identical particles. It also cannot account for the 
consequences that follow from the symmetry properties of nuclear wave functions: Localised rotational states are unphysical representations of molecules that exist in form of 
nuclear spin isomers.

The basis for our critique is the following argument: The {frozen-nuclei approximation} (implicitly) relies on a two-step mechanism. Before manipulating electrons or internal nuclear 
motions, some form of external interaction, for example an electromagnetic field, creates narrowly localised rotational states of the molecule. In the limit of infinite nuclear 
masses, common models assume, such states converge to a classical configuration with fixed nuclear coordinates,\autocite{Primas.1983,Primas.1998} which then are used in 
simulations of attosecond experiments.\autocite{Kling.2008,Haessler.2010,Salieres.2012,Spanner.2013,Diveki.2013,Vrakking.2014,Ramasesha.2016,Peng.2019} Yet, 
localised rotational states are unphysical if the rotational motion can be described in terms of permutations of  identical nuclei. Either these states cannot exist because they 
directly violate the spin-statistics theorem; or they cannot be created because they represent coherent superpositions of states belonging to different nuclear spin isomers of the 
molecule. Hence, as long as the existence of stable nuclear spin isomers of the studied molecule is a legitimate assumption---which is usually the 
case\autocite{Chapovsky.1999}---localised states are physically forbidden, and the {frozen-nuclei approximation} fails.

To unfold our critique, we begin with a systematic symmetry analysis of the two-step model that is the underlying assumption of the {frozen-nuclei approximation} by discussing the 
Molecular Symmetry (MS) groups of rigid molecules\autocite{Hougen.1962,Hougen.1963,LonguetHiggins.1963,Bunker.1998,Ezra.1982,Watson.1975} in electromagnetic fields. 
This analysis allows us to draw some general conclusions on the existence of nuclear spin isomers for rigid molecules, which eventually let us define the conditions for the 
(non-)existence of localised states. Our inference that localised rotational states, if at all, cannot be created on the time-scale of molecular control experiments leads us to the 
conclusion that the results of recent theoretical and experimental studies on molecular control have to be modified to be consistent with quantum theories. In particular, we call 
attention to the fact that adequate electronic wave functions must be (anti-)symmetrised twice: with respect to the exchange of electrons \textit{and} with respect to the 
exchange of identical nuclei.

\section{Symmetry and two-step mechanisms}
\label{sec:sym-2step}

\begin{figure}[tb!]
\centering{\includegraphics[width=0.625\textwidth]{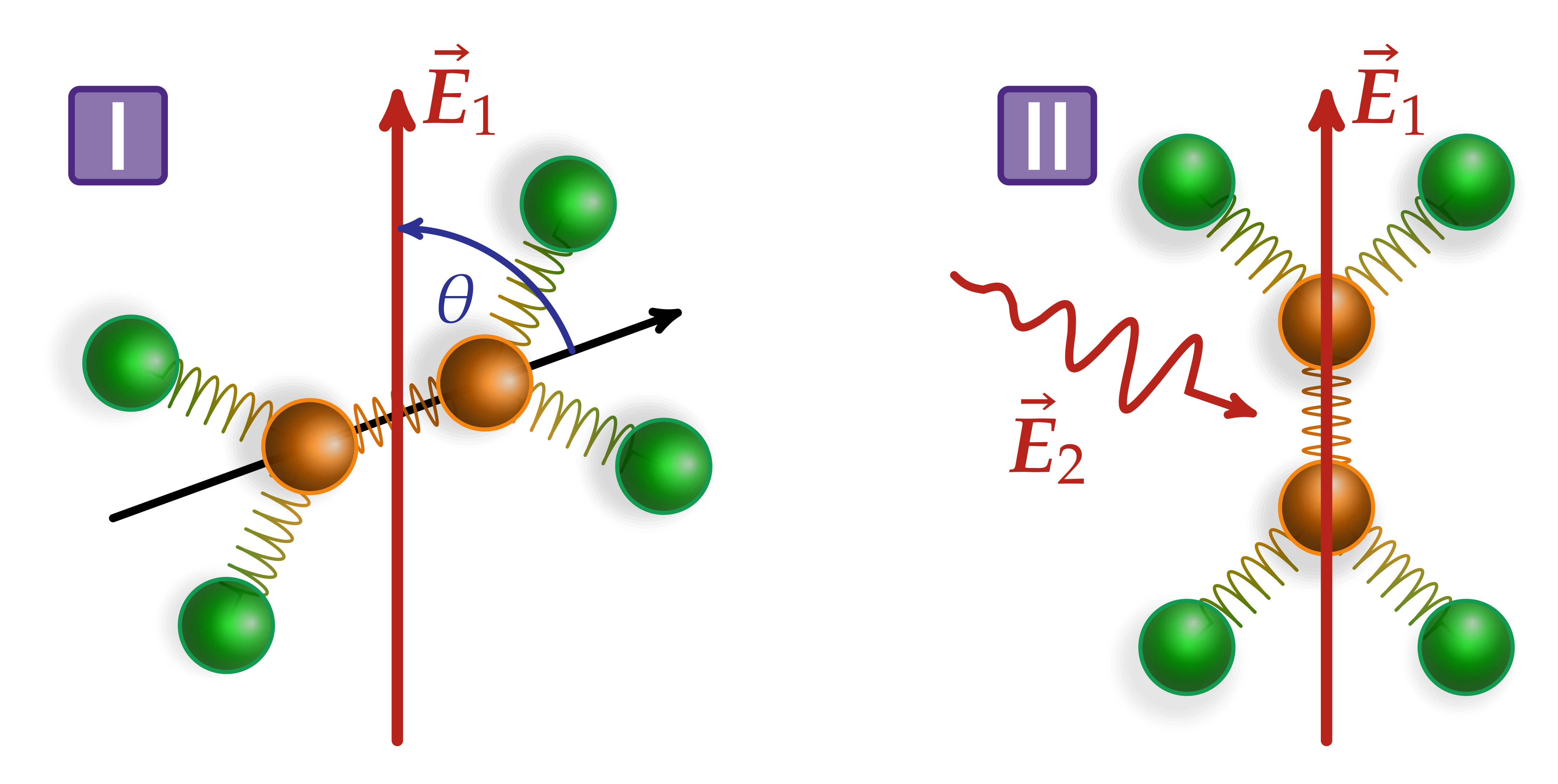}}


\caption{Cartoon of the two-step mechanism, which is often employed in molecular control communities.  \textit{Step I:} The molecule is aligned or oriented by an external 
interaction, for example created by an electromagnetic field $\bm E_1$. \textit{Step II:} The molecular motion of interest is manipulated by a second laser pulse $\bm E_2$.}
\label{Fig:two-step}
\end{figure}

One of the most prominent assumptions in studies on the control of molecular processes is the two-step model. Within this model, scientists posit that it is possible to align or 
orient the molecule they want to study along one or more of its principal axes, before the molecular process they are actually interested in is manipulated; see
\autoref{Fig:two-step} for an illustration. The need for such model is perfectly obvious: To successfully excite molecular motions with light selectively, no matter if electronic or 
nuclear, the polarisation of the external field relative to the molecular frame is often decisive. Therefore, to effectively control the motion of interest, we must be able to define 
the polarisation angle of the external field relative to the molecule. Simulations employing the frozen-nuclei approximation assume that step I in \autoref{Fig:two-step} is perfectly 
realised.

\thisfloatsetup{floatwidth=0.625\textwidth,capposition=beside}

In experiments  on molecular control, however, molecules are typically in delocalised rotational eigenstates before they interact with any laser pulse. Thus, the field $\bm E_1$ in
\autoref{Fig:two-step} must be able to create localised rotational states by superimposing delocalised rotational eigenstates of the molecule, which, in turn, allow for 
approximating the nuclear spatial distribution by classical coordinates. The rotational control with electromagnetic fields is one method that offers a route to such localised states 
by precisely steering the orientation or the alignment of molecules: The orientation of (cold) molecules can be effectively controlled with static electromagnetic fields by creating 
\textquote{field-dressed} eigenstates;\autocite{Friedrich.1991,Lemeshko.2013} employing off-resonant, femtosecond, picosecond, or nanosecond laser pulses, alignment of 
molecules is achieved by exciting rotational wave packets that are well localised along at least one of the principal axes.\autocite{Stapelfeldt.2003,Lemeshko.2013}

Yet, as we argue in \autoref{sec:frozen-nuc-localised}, localised rotational states are unphysical for molecules with identical nuclei. Basic to understanding why the 
frozen-nuclei approximation fails to describe molecules with identical nuclei is a symmetry analysis of the potential that is created by the electromagnetic fields aligning or 
orienting the molecule. In particular, the structure of the MS group\autocite{Hougen.1962,Hougen.1963,LonguetHiggins.1963,Bunker.1998} of the confined molecules is of 
central importance for our argument. Therefore, we show in the following that, for any rigid molecule in an electromagnetic field, the permutation subgroup of its MS groups, 
$\rm  G^{psms}$, can be decomposed into cyclic subgroups. Due to this partition of $\rm  G^{psms}$, we are not only able to conveniently analyse the rotational motions of a 
molecule in terms of symmetry. By using the permutation subgroup of the MS group, we can also derive general conditions for the existence of nuclear spin isomers of rigid 
molecules. We close this Section by pointing out why the theory of MS groups actually suggests that localised states are a reasonable description of confined molecules, which 
might explain the popularity of the frozen-nuclei approximation among theoreticians of molecular control.

\subsection{Confining molecules in space with electromagnetic fields}
A group of techniques that has become widespread during the last three decades is the control of molecular motions with electromagnetic fields. Two current standard approaches 
are standing to reason to realise the presumptions of the two-step model: the orientation of polar molecules with static electric fields,\autocite{Friedrich.1991,Lemeshko.2013} 
and the alignment of polarizable molecules with off-resonant, moderately intense laser pulses.\autocite{Friedrich.1995,Stapelfeldt.2003,Seideman.2005,Lemeshko.2013} Using 
these techniques, the electromagnetic field confines the molecule along its main molecular axis by exciting superpositions of field-free rotational eigenstates.

\thisfloatsetup{floatwidth=\textwidth,capposition=bottom}
\begin{figure}[tb!]
\centering{\includegraphics[width=0.925\textwidth]{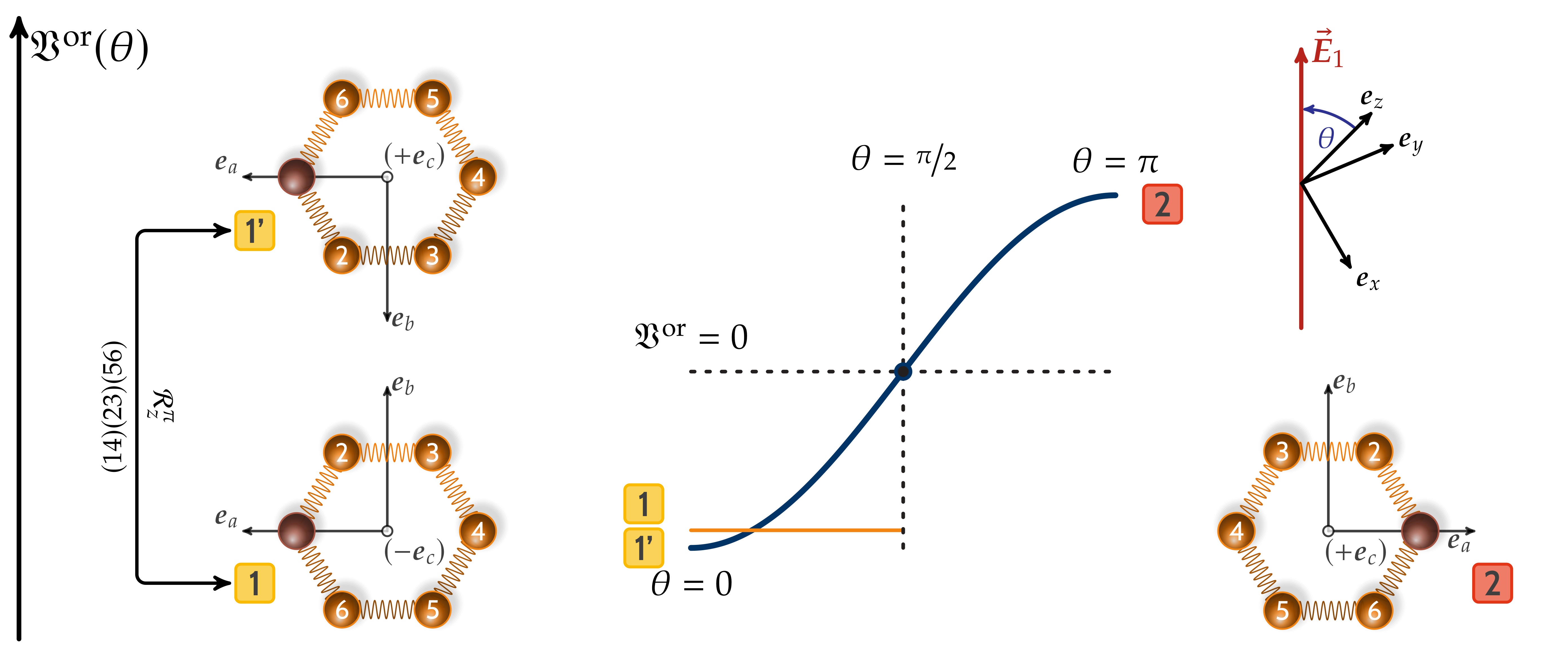}}

\vspace*{0.75\baselineskip plus 0.125\baselineskip}

\caption{The potential created by a linearly polarized orienting electric field, see \autoref{Eq:ham-or-lin}, along $\theta$ for one field strength $|\bm E|$. Due to the 
symmetry-breaking effect of the field, the aligned configuration $1$ ($\theta=0$, \textit{i.e.} ${\bm e}_Z$ and ${\bm e}_z\equiv{\bm e}_a$ are parallel) and the anti-aligned 
configuration $2$ ($\theta=\pi$, \textit{i.e.} ${\bm e}_Z$ and ${\bm e}_z\equiv{\bm e}_a$ are anti-parallel) of iodobenzene are energetically inequivalent. If the molecule 
belongs to the MS groups $\rm C_n(M)$ or $\rm C_{nv}(M)$ with $\rm n\geq 2$, the configurations $1$ and $2$ are not unique. For iodobenzene, there exist a second oriented 
version $1'$, which belongs to the same configuration in $\theta$, but which corresponds to $\chi_{1'}=\chi_1+\pi$. The configurations $1$ and $1'$ are interconverted by the 
permutation $(26)(35)$, representing the exchange (of groups) of identical nuclei. We neither show the hydrogen atoms nor the iodine of iodobenzene, because each of them is 
rigidly attached to exactly one carbon atom.\\}
\label{Fig:pot-or}
\end{figure}

For the approach using static electric fields to work, the molecule needs to be polar,\autocite{Friedrich.1991,Lemeshko.2013} \textit{i.e.} it must have a permanent dipole moment 
$\bm\mu$. Here, the field-matter Hamiltonian ${\mak H}^{\rm or}$ in the quantum-classical dipole approximation writes\autocite{Friedrich.1991,Lemeshko.2013}
\be
\label{Eq:ham-or}
{\mak H}^{\rm or}
=
-{\bm\mu}\cdotp {\bm E}\;.
\ee
and reduces to
\be
\label{Eq:ham-or-lin}
{\mak H}^{\rm or}
\equiv
{\mak V}^{\rm or}
=
-{\mu}_z\cdotp |{\bm E}|\cdotp \cos\theta
\ee
in case the electric field $\bm E$ is linearly polarized. In \autoref{Eq:ham-or-lin}, the Euler angle $\theta$ characterizes the orientation of the molecule-fixed $\bm e_z$-axis 
with respect to the space-fixed $\bm e_Z$-axis.

Because the Hamiltonian in \autoref{Eq:ham-or} and \autoref{Eq:ham-or-lin}, respectively, is time-independent, the field creates a potential ${\mak V}^{\rm or}$ that causes 
the orientation of the molecule. We can quantify its effect by solving the time-independent Schrödinger equation\autocite{Friedrich.1991}
\be
\label{Eq:tise-pen}
\left({\mak H}^{\rm rot}+{\mak V}^{\rm or}\right)\Phi^{\rm pen}(\theta,\phi,\chi) =  E^{\rm pen}\Phi^{\rm pen}(\theta, \phi,\chi)
\ee
with ${\mak H}^{\rm rot}$ being the rotational Hamiltonian of a rigid molecule. As a result, we obtain the pendular energies $E^{\rm pen}$ and the pendular states 
$\Phi^{\rm pen}$ as a function of the Euler angels $\theta, \phi,\chi$. At least for low pendular energies, pendular states are highly confined in $\theta$,\autocite{Friedrich.1991} 
and it is thus possible to orient the polar molecule along the space-fixed $\bm e_Z$-axis. In \autoref{Fig:pot-or}, we show the potential for orienting 
iodobenzene\autocite{Filsinger.2009} as an example.

If the molecule does not have a permanent dipole moment, it is still possible to confine it in space such that the premises of the two-step model are approximately true. Properly 
designed, off-resonant, moderately strong laser pulse with envelope $\bm\epsilon$ are capable of creating alignment of molecules in 
space.\autocite{Friedrich.1995,Lemeshko.2013,Stapelfeldt.2003,Seideman.2005} If the central frequency of the pulse is far detuned from any molecular transition and the optical 
cycles are much faster than a typical  (classical) rotation period of the molecule, the relevant field-matter Hamiltonian $\mak H^{\rm alg}$ reduces 
to\autocite{Stapelfeldt.2003,Seideman.2005,Moiseyev.2006}
\be
\label{Eq:ham-align}
{\mak H}^{\rm alg}
=
-\frac{1}{4}{\bm\epsilon}^{\dagger}(t)\cdotp {\bm\alpha}\cdotp{\bm\epsilon}(t)\;.
\ee
As \autoref{Eq:ham-align} shows, the field interacts with the dynamic polarizability $\bm\alpha$ of the molecule, and the only relevant part of the laser field is the envelope 
$\bm \epsilon(t)$. 

In case of a laser pulse linearly polarised in the space-fixed $\bm e_Z$-direction, or a laser pulse circularly polarised in the space-fixed $(\bm e_X,\bm e_Y)$-plane, 
\autoref{Eq:ham-align} reduces to\autocite{Artamonov.2008}
\be
\label{Eq:ham-align-lin}
{\mak H}^{\rm alg}
=
{\mak a}\frac{\epsilon^2(t)}{4}\left(\alpha^{zx}\cos^2\theta +  \alpha^{yx}\sin^2\theta\sin^2\chi\right)\;.
\ee
In \autoref{Eq:ham-align-lin}, $\mak a=\nicefrac{1}{2}$ for a circularly polarised field, and $\mak a=-1$ for a linearly polarised field, respectively; the Euler angle $\theta$ 
specifies the angle between the principal axis $\bm e_{z}$ corresponding to the largest or smallest moment of inertia and the polarisation axis of the field; the Euler angle $\chi$ 
defines the angle between the section line of the ${\bm e}_X{\bm e}_Y$-plane and ${\bm e}_x{\bm e}_y$-plane and the molecule-fixed ${\bm e}_y$-axis; and the quantities 
\addtocounter{equation}{-1}
\bse
\bea
\alpha^{zx}
&\equiv&
\alpha_ {zz}-\alpha_{xx} \\
\alpha^{yx}
&\equiv&
\alpha_ {yy}-\alpha_{xx}
\eea
\ese
are the generalized, molecule-fixed polarisability anisotropies of the molecule. Hence, for this approach to work, the molecule under investigation must have non-vanishing 
polarisability anisotropies. Solving the time-dependent Schrödinger 
\be
\label{Eq:tdse-rot}
{\rm i}\hbar\frac{\partial}{\partial t}\Psi^{\rm rot}(\theta, \phi,\chi,t) = \left({\mak H}^{\rm rot}+{\mak H}^{\rm alg}\right) \Psi^{\rm rot}(\theta, \phi,\chi,t)
\ee
makes it possible to study the effect of the interaction \autoref{Eq:ham-align-lin} on the rotational motions of the molecule.

\begin{figure}[tb!]
\centering{\includegraphics[width=0.925\textwidth]{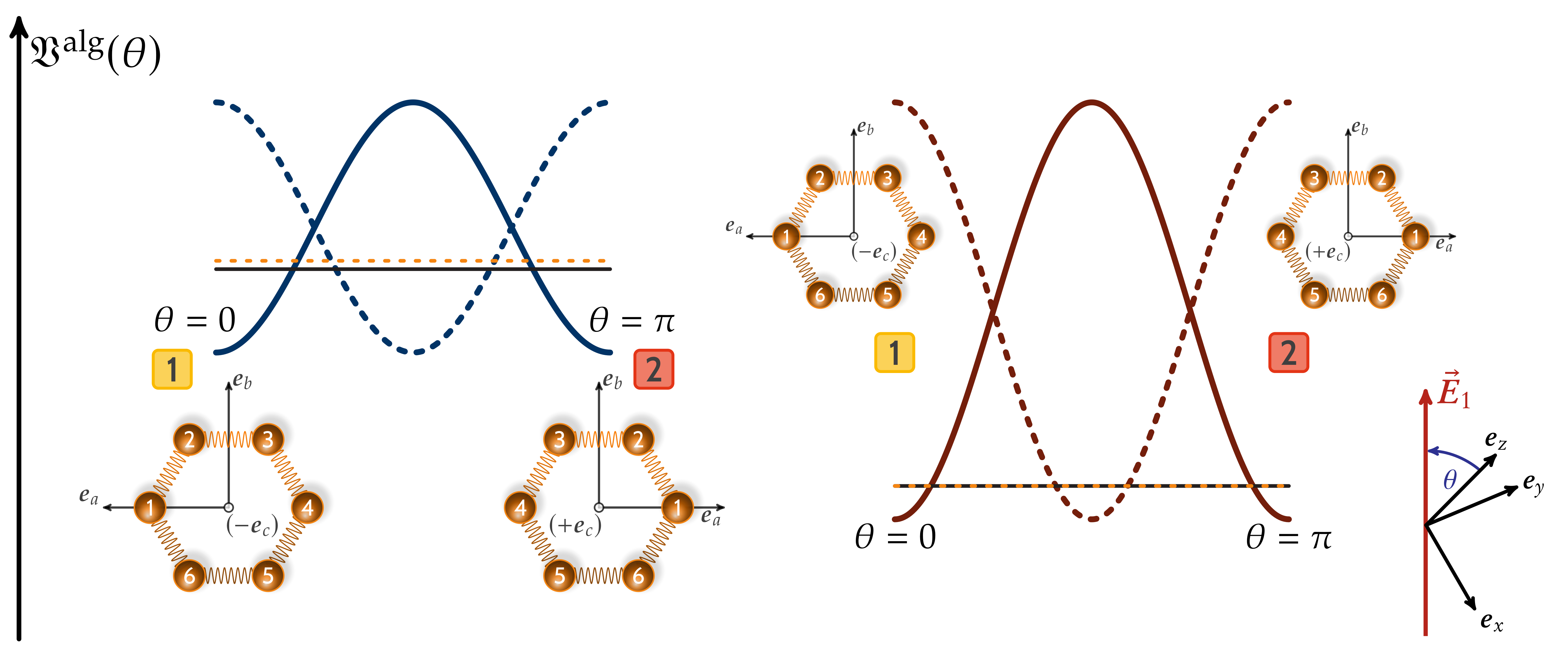}}

\vspace*{0.75\baselineskip plus 0.125\baselineskip}

\caption{The adiabatic potential created by a linear polarised (solid blue and brown lines) and circular polarised (dotted blue and brown lines) aligning laser field, see 
\autoref{Eq:ham-align-lin} and \autoref{Eq:pot-alg}, as a function of $\theta$ for two different field strengths $|\bm \epsilon|$. If the strength of the laser field becomes sufficiently 
high, tunnelling between the aligned version $1$ ($\theta=0$, \textit{i.e.} ${\bm e}_Z$ and ${\bm e}_z\equiv{\bm e}_c$ are parallel) and the anti-aligned version $2$ ($\theta=\pi$, 
\textit{i.e.} ${\bm e}_Z$ and ${\bm e}_z\equiv{\bm e}_c$ are anti-parallel) of benzene occurs on time-scales much longer than typical experiments on  molecular alignment (right 
picture). Then, (low lying) pendular energies (black solid line and dashed orange line) are structurally degenerate. If the barrier is low, tunnelling splittings become observable (left 
picture). We do not show the hydrogen atoms of benzene, because each of them is rigidly attached to exactly one carbon atom.\\}
\label{Fig:pot-aus-st}
\end{figure}

If the duration of the aligning laser pulse is much longer than a typical (classical) period of rotation, we can consider the envelope of the field to be constant over the time-scale 
of rotations. This, in turn, allows us to use the adiabatic theorem of quantum mechanics to solve \autoref{Eq:tdse-rot}. Then, the time-dependent envelope $\epsilon(t)$ in 
\autoref{Eq:ham-align-lin} can be replaced by the peak field strength $\epsilon_0$ of the laser pulse, thus making all Hamiltonians in \autoref{Eq:tdse-rot} time-independent. 
The field-free rotational states then adiabatically evolve into pendular states $\Phi^{\rm pen}$, similar to the case of orienting molecules with static 
fields.\autocite{Friedrich.1991,Friedrich.1995} Pendular states in the context of alignment are the solution of the time-independent Schrödinger equation \autoref{Eq:tise-pen} 
with $\mak V^{\rm or}$ being replaced by 
\be
\label{Eq:pot-alg}
{\mak V}^{\rm alg}
\equiv
{\mak a}\frac{|\epsilon_0|^2}{4}\left(\alpha^{zx}\cos^2\theta +  \alpha^{yx}\sin^2\theta\sin^2\chi\right)\;.
\ee
Consequently, the eigenstates of the \textquote{field-dressed} Hamiltonian ${\mak H}^{\rm rot}+{\mak V}^{\rm alg}$ can fully describe the alignment that is assumed to be 
perfectly realised within the two-step model. \Autoref{Fig:pot-aus-st} shows the potential \autoref{Eq:pot-alg} for benzene for two different field strengths.

It is an interesting feature of non-resonant laser pulses that this method of molecular alignment can be applied to polar and non-polar molecules alike. Due to the rapid 
oscillations of the laser field compared to the rotational motions, the interaction term including the permanent dipole moment of the molecule is on average zero. Hence, the 
effective interaction between laser pulse and molecule is the same, no matter if the molecule possesses a permanent dipole moment or not. In this sense, the method of 
non-resonant laser pulses is more general than the approach of orienting molecules with static electromagnetic fields. As we see in the following, however, there are subtle 
differences in terms of symmetry, leading to different limitations of the frozen-nuclei approximation. 

\thisfloatsetup{floatwidth=\textwidth,capposition=bottom}

\subsection{MS groups and properties of rigid molecules}
To understand why the frozen-nuclei approximation needs to be discarded for molecules with identical nuclei, analysing the symmetry of the rotational Hamilton 
${\mak H}^{\rm rot}$ dressed with the potentials \autoref{Eq:ham-or-lin} and \autoref{Eq:pot-alg}, respectively, is crucial. A systematic and consistent approach to the 
symmetry of molecules is offered by the theory of MS groups.\autocite{Hougen.1962,Hougen.1963,LonguetHiggins.1963,Bunker.1998,Bunker.2005} The MS 
group is the set of all \textquote{feasible} permutations $\mal P$ and permutation-inversions $\mal P^*$ of identical nuclei,\autocite{LonguetHiggins.1963} \textit{i.e.} those 
$\mal P$ and $\mal P^*$ that interconvert versions of the molecule that are not separated by insuperable energy barriers;\autocite{Bunker.1998,Bunker.2005} see in particular 
Ref. \citenum{Bunker.1998} for a systematic introduction to the theory of MS groups.

The field-free Hamiltonian in the electromagnetic approximation is invariant in the MS group, so is the Hamiltonian ${\mak H}^{\rm rot}$.\autocite{Bunker.1998,Bunker.2005}
Including the field-matter interactions \autoref{Eq:ham-or} and \autoref{Eq:ham-align}, however, might break some symmetries of the field-free 
molecule.\autocite{Watson.1975,Wilson.1980,Bunker.1998} From group theoretical 
considerations,\autocite{Watson.1975,Wilson.1980,Bunker.2005,Grohmann.2011,Grohmann.2017,Grohmann.2018c,Grohmann.2020} it follows that ${\mak H}^{\rm alg}$ has to be 
invariant in the MS group, too, while for ${\mak H}^{\rm or}$ only permutations $\mal P$ of identical nuclei remain symmetry operations. Thus, for both types of interactions, the 
set of all feasible permutations $\mal P$ is a symmetry group of the molecule in the electromagnetic field. This symmetry group is called the permutation subgroup of the MS 
group, $\rm G^{psms}$.\autocite{Watson.1975,Grohmann.2011,Grohmann.2017,Grohmann.2018c} 

For rigid molecules, which are often studied in attosecond research, a general analysis of the MS group and its permutation subgroup is possible. The MS group of a rigid 
molecule is always isomorphic to the molecular point group of its equilibrium structure;\autocite{Hougen.1962,Hougen.1963,Bunker.1998,Steinborn.1993} the 
\textquote{quantum effects} due to large amplitude contortions are negligible on the time-scale of the experimental technique that is used to characterize the 
molecule.\autocite{Bunker.1998} Due to this isomorphism, we are able to associate a classical structure to a molecule,\autocite{Primas.1983,Primas.1985,Primas.1990b} which is 
defined by the nuclear coordinates at the global minimum of the \textsc{Born-Oppenheimer} potential energy surface. With the help of the irreducible representations of its (E)MS 
group, we are able to uniquely classify any state of a molecule by symmetry labels.\autocite{Wigner.1959,Wilson.1980,Papousek.1982,Bunker.1998,Bunker.2005,Atkins.2010} 

In molecular control simulations, the rotations of a molecule are often treated within the rigid rotor 
model.\autocite{Wilson.1980,Papousek.1982,Bunker.1998,Bunker.2005,Atkins.2010} Here, each type of rigid rotor can only belong to a specific class of MS groups, be it a 
linear rotor, a symmetric top, a asymmetric top, or a spherical top. Spherical tops, for instance, have to belong to MS groups that are isomorphic to point groups containing rotations 
that are generated by at least two different $\mal C_3$ axes.\autocite{Bunker.1998,Bunker.2005} Two examples for such groups are $\rm T_d(M)$ or $\rm O_h(M)$. Using the 
results of representation theory,\autocite{Tinkham.2003,Heine.1993,McWeeny.2002,Bunker.1998} we are able to deduce that spherical tops neither have a permanent dipole 
moment, nor non-vanishing polarizability anisotropies. Thus, both types of interactions, \autoref{Eq:ham-or} and \autoref{Eq:ham-align}, are not capable of confining spherical tops. 
Consequently, spherical tops are not relevant for our discussion.

All other types of rigid rotors, however, can have non-zero polarizability anisotropies or permanent dipole moments, and we can identify the type of rotor and non-vanishing 
molecular properties by knowing the MS group of the molecule. For our discussion, the following symmetry properties of rigid rotors are 
relevant:\autocite{Bunker.1998,Bunker.2005}

\bit
[leftmargin=7.5em,
labelwidth=7.25em,
labelsep=0.25em,
rightmargin=0pt,
topsep=0.25\baselineskip,
itemsep=0.5\baselineskip,
format={\itshape\color{DFG-blau}},
align=left
]
\item[Symmetric tops]
Two of the three principal moments of inertia and moments of polarizability are identical. One distinguishes oblate symmetric tops with $I_{\rm a}=I_{\rm b}<I_{\rm c}$ from 
prolate symmetric tops with $I_{\rm a}<I_{\rm b}=I_{\rm c}$.  A necessary condition for a rigid molecule to be a symmetric top is that among the symmetry elements generating 
the operations of its point group is exactly one ${\mal C}_{\rm n}$ with ${\rm n}\geq 3$. Hence, molecules are symmetric tops if they belong to the groups $\rm C_n(M)$, 
$\rm S_n(M)$, $\rm D_n(M)$, $\rm C_{nh}(M)$, $\rm C_{nv}(M)$, $\rm D_{nh}(M)$ and $\rm D_{nd}(M)$ with $\rm n\geq 3$. Moreover, all molecules with symmetry group 
$\rm D_{2d}(M)$ are symmetric tops as well. For any symmetric top, the second term on the right hand side of \autoref{Eq:ham-align-lin} and \autoref{Eq:pot-alg} vanishes, because 
the two principal moments of the polarizability $\alpha_{xx}$ and $\alpha_{yy}$ are identical. Furthermore, only if the symmetric top belongs to the groups $\rm C_{n}(M)$ or 
$\rm C_{nv}(M)$, it has a non-vanishing permanent dipole moment. For these molecules, an interaction of the type \autoref{Eq:ham-or} is capable of confining them.

\item[Asymmetric tops]
All three moments of inertia and principal polarizabilities are different from each other. A necessary condition for a rigid molecule to be an asymmetric top is that among the 
symmetry elements generating the operations of its point group are only $\text{C}_2$ axes or no symmetry axis. Consequently, molecules that belong to the groups 
$\rm C_1(M)$, $\rm C_s(M)$, $\rm C_i(M)$, $\rm C_2 (M)$, $\rm D_2(M)$, $\rm C_{2h}(M)$, $\rm C_{2v}(M)$, and $\rm D_{2h}(M)$ are asymmetric tops. For this type of molecule, 
none of the terms on the right hand side of \autoref{Eq:ham-align-lin} and \autoref{Eq:pot-alg} is zero because of molecular symmetry. If the asymmetric top belongs to the MS 
groups $\rm C_1(M)$, $\rm C_s(M)$, $\rm C_{2}(M)$ or $\rm C_{2v}(M)$, it also has a non-zero permanent dipole moment, and orienting the molecule is possible by using an 
interaction like \autoref{Eq:ham-or}. Molecules with symmetry $\rm C_1(M)$, $\rm C_i(M)$, or $\rm C_s(M)$ are irrelevant for our discussion, because they do not contain 
permutations of identical nuclei except the identity.\autocite{Hougen.1962,Hougen.1963,Bunker.1998}

\item[Linear molecules]
Two of the three principal moments of inertia and polarizability are identical, but $I_{\text{a}}\ll I_{\text{b}}= I_{\text{c}}$. Hence, the rotation about the bond axis cannot be 
excited without the molecule being destroyed. Linear molecules either belong to the (E)MS groups $\rm C_{\infty v}(M)$ or $\rm D_{\infty h}(M)$. The Hamiltonian for the 
field-matter interaction in \autoref{Eq:ham-align-lin} and \autoref{Eq:pot-alg} is the same as for a symmetric top. For our discussion, only molecules with feasible permutations of 
identical nuclei are relevant; they necessarily belong to $\rm D_{\infty h}(M)$ with permutational subgroup $\rm C_2(M)$.\autocite{Bunker.1998} These type of molecules do not 
have a permanent dipole moment. Moreover, it is legitimate to consider linear molecules as a special case of symmetric top molecules, as the following discussion shows.
\eit

Summarizing, the different types of rigid rotors can be sub-classified according to their MS group. This allows us to make some general conclusions about the possibility of 
applying the frozen-nuclei approximation, as we show in the following: If the MS group contains pure permutations that are different from the identity, the frozen-nuclei 
approximation is physically wrong and needs to be discarded.

\subsection{Confinement and symmetry: symmetric tops}
\label{subsec:perm-sym-sym-top}

Central to our argument is the structure of the MS group of the molecule that is manipulated. In particular, it is imperative that, first, we are able to decompose the MS group into 
its permutation subgroup, $\rm G^{psms}$, and one other subgroup of order 2; and second, we can write $\rm G^{psms}$ as product of cyclic subgroups. As we show in 
\autoref{sec:frozen-nuc-localised}, the general structure of MS groups for rigid molecules has strong implications for the validity of the frozen-nuclei approximation. Yet, some of 
the results we present in the following are not only valid for rigid molecules but for molecules in general, as \autoref{sec:append}, \autoref{subsec:types-ms} points out.

\thisfloatsetup{floatwidth=0.65\textwidth,capposition=side}
\begin{figure}[tb!]
\centering{\includegraphics[width=0.625\textwidth]{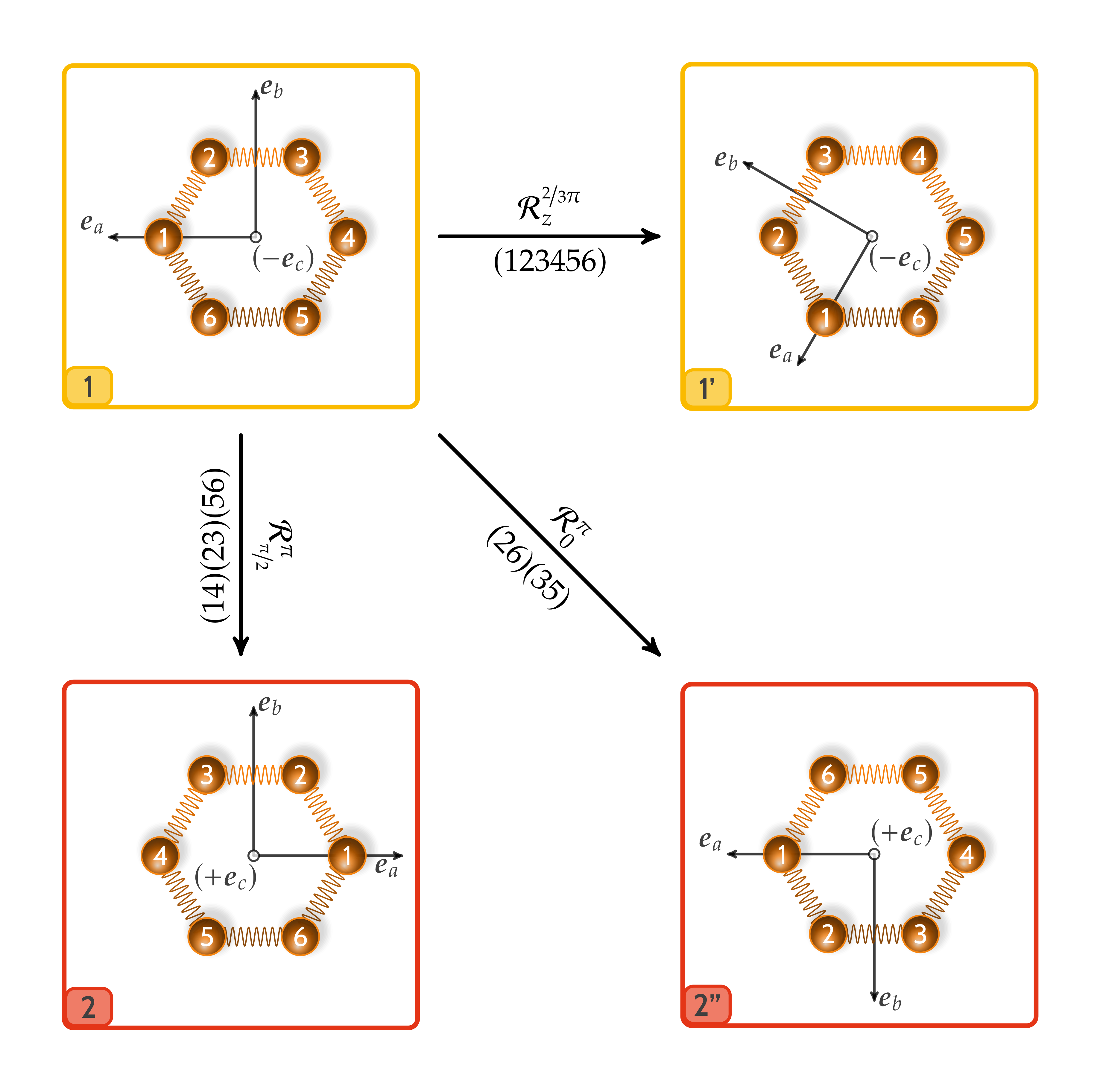}}


\caption{Four classical configurations of the nuclear frame of benzene, two corresponding to the aligned potential-dressed version (upper row: $1$,$1^{\prime}$) and two 
representing the anti-aligned potential-dressed version (lower row: $2$, $2^{\prime\prime}$) of benzene. All configurations can be generated out of version 1 by permuting 
identical nuclei. We do not show the hydrogen nuclei; $(12 ...)$ means carbon nucleus $1$ with hydrogen nucleus $1$ is replaced by carbon nucleus $2$ with hydrogen nucleus $2$, 
and so on. Applying the frozen-nuclei approximation is equivalent to choosing one of all possible configurations.}
\label{Fig:version-benzene}
\end{figure}

\paragraph{An illustrative example: aligning benzene}
Let us consider symmetric top molecules first, using the alignment of benzene as an example. Here, the potential created by the aligning laser pulse is only dependent on 
$\theta$; see \autoref{Eq:ham-align-lin} for $\alpha_{xx}$ $=$ $\alpha_{yy}$ and \autoref{Fig:pot-aus-st} for a graphical illustration. As the potential has two equivalent 
minima with identical energy, two \textquote{potential-dressed versions} of benzene exist, one corresponding to the aligned configuration ($\theta=0$, \textit{i.e.} 
${\bm e}_Z$ and ${\bm e}_z\equiv{\bm e}_c$ are parallel) of benzene and one belonging to its anti-aligned version ($\theta=\pi$, \textit{i.e.} ${\bm e}_Z$ and 
${\bm e}_z\equiv{\bm e}_c$ are anti-parallel).

In the field-free theory of MS groups, different versions of molecules are defined as distinct ways of labelling the identical nuclei of a rigid structure that cannot be mapped onto 
each other by a rigid rotation. Recently, we have shown that the potential-dressed versions can be understood similarly if we take into account that, in the presence of a 
electromagnetic field, the isotropy of space is broken.\autocite{Grohmann.2020} If the external potential that is created by the field has minima that are interconverted by 
permutations, permutation-inversions and/or elements of the spatial rotation group $\mal K^{\rm spa}$, they belong to different physical situations that are separated by 
potential barriers. This justifies to call them different---potential-dressed---versions of the molecule. As for the versions of molecules in the field-free theory of MS groups, certain 
presumptions have to be justified to define them properly.\autocite{Grohmann.2020,Grohmann.2021b}

For symmetric tops, however, to each potential-dressed version there exist more physically equivalent nuclear configurations than just the two that \autoref{Fig:pot-aus-st} 
shows. As \autoref{Fig:version-benzene} illustrates, the two potential-dressed versions are not unique: The permutation $(123456)$, for example, changes version to $1$ to $1'$, 
for which the configuration in $\theta$ is the same, but $\chi$ is changed to $\chi+\nicefrac{\pi}{3}$. Since $\mak V^{\rm alg}$ is independent of $\chi$, configurations with 
primes are not separated by potential barriers, and the pulse cannot control the motion in $\chi$. Yet, to each primed and non-primed configuration belongs a different set of 
rotational coordinates, which represent different classical nuclear configurations of the molecule. Applying the frozen-nuclei approximation, and thus fixing all nuclear coordinates, 
is identical to single out one of these configurations.

Using the MS group $\rm D_{6h} (M)$ of benzene allows for a systematic analysis of all equivalent configurations. This group decomposes according to
\be
\label{Eq:D-6h}
{\rm D_{6h}(M)}
=
{\rm D_6(M)} \otimes 
\left\{{\rm E},{\rm E^*} \right\}
=
{\rm D_6(M)} \otimes {\rm C_s(M)} \;,
\ee
where $\rm C_s(M)$ is isomorphic to the point group $\rm C_s$; see also \autoref{sec:append}, \autoref{subsec:types-ms} for more explanations on the relation of point group 
operations and operations of the MS group. The group
\be
{\rm G^{psms}}[{\rm D_{6h}(M)}]={\rm D_6(M)}
\ee
defines the permutation subgroup of $\rm D_{6h} (M)$; it contains all feasible permutations of identical nuclei and, for benzene in its equilibrium structure, it is isomorphic to the 
point group $\rm D_6$.\autocite{Bunker.1998} For our later deliberations, it is interesting to note that 
\bse
\addtocounter{equation}{-1}
\be
\label{Eq:D-6-sp} 
{\rm D_6(M)} = {\rm C_6(M)} \rtimes {\rm C^{\perp}_2(M)} 
\ee
with $\rtimes$ defining the semi-direct product of the groups ${\rm C_6(M)}$ and $\rm C^{\perp}_2(M)$. Both ${\rm C_6(M)}$ and $\rm C^{\perp}_2(M)$ are cyclic and their 
presentations can be written as\autocite{Mirman.2007}
\bea
\label{Eq:C-6} 
{\rm C_6(M)}
&=&
\big<g=(123456) \,\big|\, g^6={\rm E} \big>
\\
\label{Eq:C-2-perp} 
{\rm C^{\perp}_2(M)} 
&=&
\big<g_{\perp}=(23)(56)\,\big|\,g_{\perp}^2={\rm E} \big> \;.
\eea
\ese
The irreducible representations of $\rm D_{6h}(M)$, ${\rm C_6(M)}$, and ${\rm C^{\perp}_2(M)}$ are all known and tabulated in the 
literature.\autocite{Altmann.1994,Bunker.1998}

For rigid molecules, any operation of the MS group, ${\mal O}^{\rm ms}$, can be formally expressed as\autocite{Bunker.1998}\nolinebreak[4] 
\be
\label{Eq:op-MS}
{\mal O}^{\rm ms}
= 
{\mal R}^{\rm eq}
\cdotp
{\mal O}^{\rm pg}
\cdotp
{\tt p}_{..ijk..}
\ee
In \autoref{Eq:op-MS}, ${\mal R}^{\rm eq}$ is an equivalent rotation, solely acting on the rotational degrees of freedom;  ${\mal O}^{\rm pg}$ is an operation of the molecular 
point group, only affecting the internal nuclear coordinates of a molecule; and  ${\tt p}_{..ijk..}$ means a relabelling of the spin variables 
$...,\Sigma_{i},\Sigma_{j},\Sigma_{k},...$ of the identical nuclei alone.\autocite{Hougen.1962,Hougen.1963,Bunker.1998} If the molecule is non-rigid, \autoref{Eq:op-MS} needs to 
be extended.\autocite{Bunker.1998,LonguetHiggins.1963,Watson.1965} In \autoref{sec:append}, \autoref{subsec:oper-ms}, we comment briefly on the rules of how to map 
${\mal O}^{\rm ms}$ on the familiar point group operations ${\mal O}^{\rm pg}$. Rigid molecules are special in so far as each one of the operations ${\mal O}^{\rm ms}$ can be 
mapped exactly on one operation of the point group, ${\mal O}^{\rm pg}$, and \textit{vice versa}.\autocite{Hougen.1962,Hougen.1963,Bunker.1998} 

To understand why the frozen-nuclei approximation is physically wrong, only the equivalent rotations in \autoref{Eq:op-MS} are relevant. For the elements of the permutational 
subgroup, $\rm G^{psms}$, of a rigid molecule, each of the permutations can be mapped on a different equivalent rotation.\autocite{Hougen.1962,Hougen.1963} Principally, two 
different types of rotations exist, $\mal R^{\pi}_{\alpha}$ and $\mal R^{\beta}_{z}$. Their meaning is the following: $\mal R^{\beta}_z$ is a rotation through $\beta$ radians about 
the molecule-fixed $\bm e_z$-axis; $\mal R^{\pi}_{\alpha}$ is a rotation through $\pi$ radians about an axis in the $\bm e_x\bm e_y$-plane making an angle $\alpha$ with the $\bm 
e_x$-axis.\autocite{Bunker.1998} Both types of equivalent rotations change the set of Euler angles $(\theta,\phi,\chi)$ according to\autocite{Bunker.1998}
\bse
\bea
\label{Eq:R-eq-alph}
\mal R^{\pi}_{\alpha}
&:& 
(\theta,\phi,\chi)\; \rightarrow\; (\pi-\theta, \phi+\pi,2\pi-2\alpha-\chi)\\
\label{Eq:R-eq-beta}
\mal R^{\beta}_{z} 
&:&
(\theta,\phi,\chi)\; \rightarrow\; (\theta, \phi,\chi + \beta)\;.
\eea
\ese

Hence, each of the permutations in \autoref{Eq:C-6} and \autoref{Eq:C-2-perp} changes the rotational configuration of the molecule in the laser field according to 
\autoref{Eq:R-eq-alph}, or \autoref{Eq:R-eq-beta}. The operations of the group $\rm C_6(M)$ correspond to rotations $\mal R^{\beta}_z$ about the molecule-fixed $\bm e_z$ axis. 
Considering $(123456)$, for example, ${\mal R^{\rm eq}}=\mal R^{\nicefrac{\pi}{3}}_z$. Hence, the operations of $\rm C_6(M)$ do not change the potential-dressed version $1$ or 
$2$, but their superscript; say $2^{\prime}\rightarrow 2^{\prime\prime}$ if we choose again $(123456)$ as an example. The operations of ${\rm C^{\perp}_2(M)}$, however, change 
the potential-dressed version of benzene, because for $(14)(23)(56)$, ${\mal R^{\rm eq}}=\mal R^{\pi}_{\nicefrac{\pi}{2}}$; see \autoref{Fig:version-benzene} for an illustration.

If we write out the direct product, \autoref{Eq:D-6-sp}, it turns out that twelve different configurations of benzene exist.\autocite{Bunker.1998} Six of them correspond to the 
aligned version of benzene, $1, \hdots , 1^{\prime\prime\prime\prime\prime}$, and six of them belong to the anti-aligned version of benzene, 
$2, \hdots, 2^{\prime\prime\prime\prime\prime}$. If scientists apply the frozen-nuclei approximation, they choose only one of these twelve configurations for their theoretical 
description. This would not be much of a problem if a localised state, being well defined at one of the twelve equivalent configurations, could be prepared by some form of 
interaction. Yet, as \autoref{sec:frozen-nuc-localised} shows, such state cannot exist.

\thisfloatsetup{floatwidth=0.525\textwidth,capposition=beside}
\begin{figure}[tb!]
\centering{\includegraphics[width=0.55\textwidth]{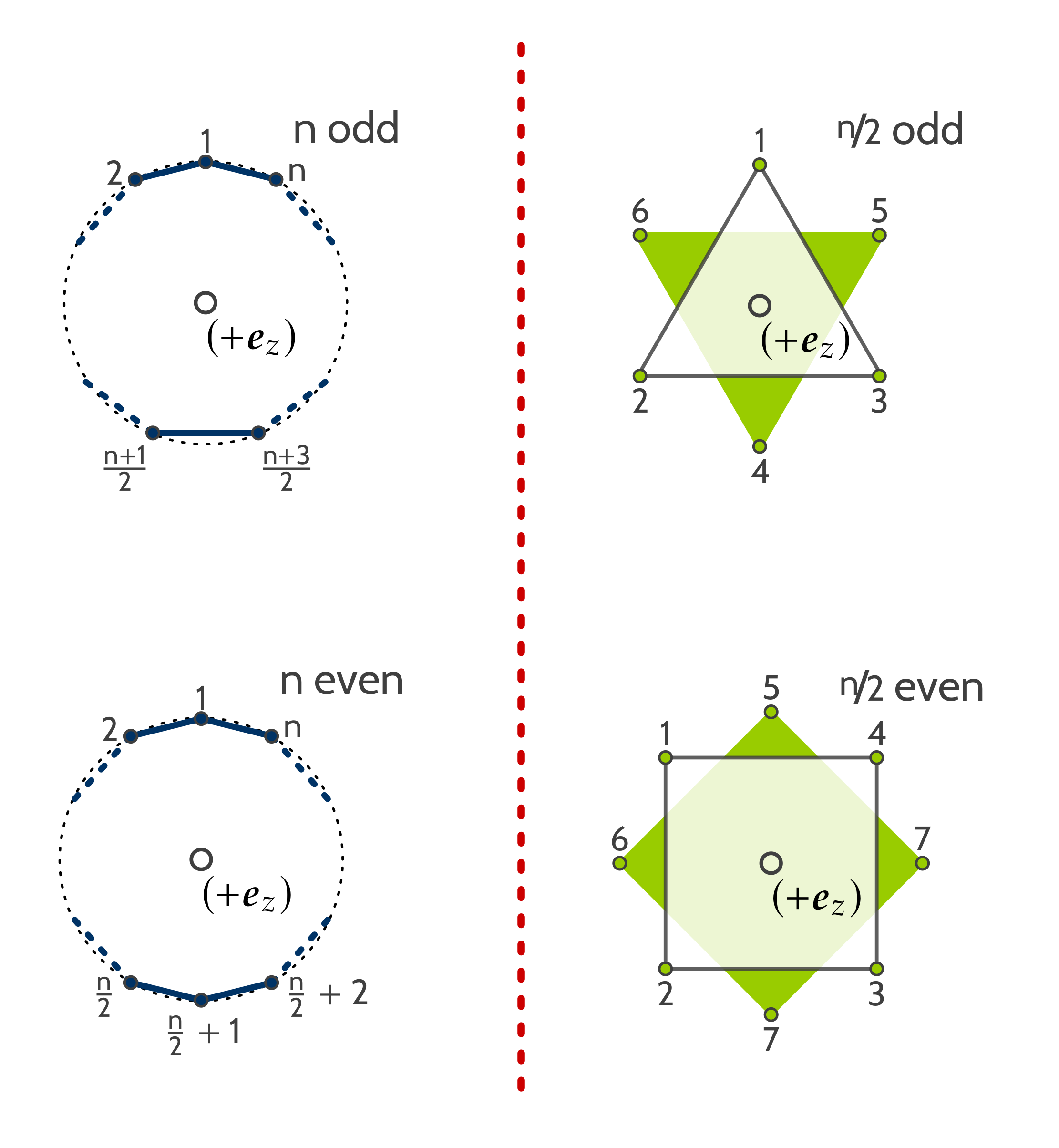}}


\caption{Geometrical realisations of the MS groups of rigid molecules. \textit{Left column:} The groups $\rm C_n(M)$ are generated by cyclic permutations reflecting the 
rotation about the main principal axis $\bm e_z$. If we allow for inversions, we obtain the groups $\rm C_{nh}(M)$. If we allow for permutation-inversions that correspond to 
reflections at planes including the main principal axis, we arrive at $\rm C_{nv}(M)$. If we allow for operations that can be mapped on rotations perpendicular to the main 
principal axis, we obtain the groups $\rm D_n(M)$ (without inversions) and $\rm D_{nh}(M)$ (with inversions). \textit{Right column:} The group $\rm S_{2n}(M)$ is generated by 
the simultaneous cyclic permutation-inversion representing the improper rotation of the corners of two staggered regular polygons about the main principal axis $\bm e_z$. If 
we further allow for permutations that can be mapped on rotations perpendicular to the main principal axis, we obtain the groups $\rm D_{nd}(M)$. The labels $1$ ... $\rm n$ 
represent identical nuclei or identical groups of nuclei.}
\label{Fig:gruppe-abstr}
\end{figure}

\paragraph{Generalizations: polarizable symmetric tops}
The results for benzene can be generalized to any rigid symmetric top if we take into account the general structure of point groups. Using the results from 
literature,\autocite{Ezra.1982,Bunker.1998} we can divide the MS groups of molecules into three types: 

\setlength\mathindent{8.25em}
\newtagform{brackets4}[\color{DFG-blau}]{\hspace*{4em}\textcolor{DFG-blau}{Eq.{\tiny\,}}{}}{{}}
\usetagform{brackets4}

\bit
[leftmargin=4em,
topsep=0\baselineskip,
itemsep=0.5\baselineskip,
format={\it\color{DFG-blau}},
labelwidth=3.75em,
align=left
]
\item[{\color{DFG-blau} Type-I}] 
MS groups that contain only permutations.

\item[{\color{DFG-blau} Type-II$^a$}] 
MS groups that contain permutation-inversions and can be written as
\bse
\be
\label{Eq:MS-typ-IIa}
{\rm G^{ms}_{II^a}}
=
{\rm G^{pdms}} \otimes \{{\rm E, \rm E^*}\}\,.
\ee 

\item[\color{DFG-blau} Type-II$^b$] 
MS groups that contain permutation-inversions and for which
\be
\label{Eq:MS-typ-IIb}
{\rm G^{ms}_{II^b}}
\neq
{\rm G^{pdms}} \otimes \{{\rm E, \rm E^*}\}\,.
\ee
\ese
\eit
\setlength\mathindent{4.25em}
\usetagform{brackets2}

See also \autoref{sec:append}, \autoref{subsec:types-ms} for more information on this typology of MS groups. 

MS groups that are isomorphic to the point groups $\rm C_n$ and $\rm D_n$ are MS groups of \textit{type-I}.\autocite{Ezra.1982,Bunker.1998} Because they only contain 
permutations, the permutational subgroup $\rm G^{\rm psms}$ is the improper subgroup of the MS group; they are identical. One presentation of the MS group $\rm C_n(M)$ 
is\autocite{Mirman.2007}
\be
\label{Eq:pres-Cn}
{\rm C_n(M)}
=
\big<g=(1,2,...,{\rm n}) \,\big|\, g^{\rm n}={\rm E} \big>\,,
\ee
and it can be understood as the group of simultaneous cyclic permutations of the corners $1,2,...,{\rm n}$ of a regular $\rm n$-sided polygon; see \autoref{Fig:gruppe-abstr} for an 
illustration.\autocite{Mirman.2007} The group $\rm D_n(M)$ writes
\be
\label{Eq:Dn-dir}
{\rm D_n(M)}
=
{\rm C_n(M)}
\rtimes
{\rm C^{\perp}_2(M)}\,,
\ee
whereas the presentation of $\rm C_n(M)$ shows \autoref{Eq:pres-Cn} and ${\rm C^{\perp}_2(M)}$ can be written as
\bse
\begin{align}
\label{Eq:pres-C2-perp-1}
{\rm C_2^{\perp}(M)}
&=
\big<g_{\perp}=(2,{\rm n})(3,{\rm n}-1)...(\nicefrac{{\rm n}}{2},\nicefrac{{\rm n}}{2}+2) \,\big|\, g^{\rm n}_{\perp}={\rm E} \big> 
&&
{\text{if {\rm n} is even}}\\
\label{Eq:pres-C2-perp-2}
{\rm C_2^{\perp}(M)}
&=
\big<g_{\perp}=(2,{\rm n})(3,{\rm n}-1)...(\nicefrac{{\rm n}+1}{2},\nicefrac{{\rm n}+1}{2}+1) \,\big|\, g^{\rm n}_{\perp}={\rm E} \big> 
&&
{\text{if {\rm n} is odd.}}
\end{align}
\ese
The labels in \autoref{Eq:pres-C2-perp-1} and \autoref{Eq:pres-C2-perp-2} have the same meaning as in the case of $\rm C_{n}(M)$; they enumerate the corners $1,2,...,{\rm n}$ of a 
regular $\rm n$-sided polygon; see also \autoref{Fig:gruppe-abstr}.\autocite{Mirman.2007}

By convention, it is the $\bm e_z$-axis about which the molecule rotates if the permutations $(1,2,...,{\rm n})$ from \autoref{Eq:pres-Cn} are applied to the (set of) identical nuclei 
$1$, $2$, ... $\rm n$.\autocite{Altmann.1994,Tinkham.2003,Cotton.1971,Heine.1993,Bunker.1998} For symmetric tops, the choice of the main principal axis is unique; by definition, 
there exists only one principal axis that reflects the permutations $(1,2,...,{\rm n})$ with ${\rm n}\geq 3$.\autocite{Bunker.1998} Consequently, the operations $(1,2,...,{\rm n})$ do 
not change the orientation of the molecule-fixed $\bm e_z$-axis, and hence, the orientation of the molecule with respect to space-fixed $\bm e_Z$-axis. If applied $k$-times, they 
do change, however, the configuration in $\chi$ according to\autocite{Bunker.1998}
\be
\label{Eq:chi-transf}
\chi 
\rightarrow
\chi + \frac{2\pi}{{\rm n}} k \;.
\ee

As \autoref{Eq:chi-transf} shows, the operations of $\rm C_n(M)$ rotate one primed classical configuration to a different primed configuration, but they do not change the 
potential-dressed version from $1$ to $2$ or from $2$ to $1$. By contrast, in case the molecule belongs to the MS group $\rm D_n (M)$, the operations from 
\autoref{Eq:pres-C2-perp-1} and \autoref{Eq:pres-C2-perp-2} do change the potential-dressed version from $1$ to $2$ or the other way around. Consequently, for molecules with 
$\rm C_n (M)$-symmetry the operations of the permutation subgroup $\rm G^{psms}$ only change the configuration in $\chi$, while for molecules with $\rm D_{n}(M)$-symmetry 
the operations of $\rm G^{psms}$ change the configuration in $\chi$ and in $\theta$. In both cases, classical molecular structures are interconverted by the permutation of (groups 
of) identical nuclei.

Lets turn to the case of symmetric top molecules with a MS group of \textit{type-II}$^a$. In this category fall the groups $\rm C_{nh}(M)$ and $\rm D_{nh}(M)$, because they can be 
written as\autocite{Ezra.1982,Altmann.1994}
\bse
\bea
{\rm C_{nh}(M)}
&=&
{\rm C_n(M)} \otimes \left\{{\rm E},{\rm E}^*\right\}
\\
{\rm D_{nh}(M)}
&=&
{\rm D_n(M)} \otimes \left\{{\rm E},{\rm E}^*\right\}\;.
\eea
\ese
Here, the groups ${\rm C_n(M)}$ and ${\rm D_n(M)}$ are identical to the permutation subgroups of ${\rm C_{nh}(M)}$ and ${\rm D_{nh}(M)}$, respectively. Consequently, 
considering the permutation subgroup $\rm G^{psms}$ of the MS group only, the same conclusion as for molecules with MS groups ${\rm C_n(M)}$ and ${\rm D_n(M)}$ apply to 
molecules with $\rm C_{nh}(M)$ and $\rm D_{nh}(M)$ symmetry, respectively. 

The structure of the groups $\rm S_n(M)$, $\rm C_{nv}(M)$ and $\rm D_{nd}(M)$ is more difficult. They belong to MS groups of \textit{type-II}$^b$ and cannot be written as a direct 
product of their permutation subgroup and the inversion group, like \autoref{Eq:MS-typ-IIa}. Yet, these groups decompose according to\autocite{Ezra.1982,Altmann.1994}
\bse
\bea
\label{Eq:Cnv-prod}
{\rm C_{nv}(M)}
&=&
{\rm C_n(M)}
\rtimes
{\rm C^{\perp}_s(M)}
\\
\label{Eq:Dnd-prod}
{\rm D_{nd}(M)}
&=&
{\rm D_n(M)}
\rtimes
{\rm C^{\perp}_s(M)}
\\
\label{Eq:Sn-prod}
{\rm S_{2n}(M)}
&=&
{\rm C_{n}(M)}
\rtimes
{\rm C_i(M)}
\quad \text{if $\rm n$ is odd.}
\eea
\ese
In all cases, \autoref{Eq:Cnv-prod}, \autoref{Eq:Dnd-prod} and \autoref{Eq:Sn-prod}, the permutation subgroups are, again, isomorphic to the groups $\rm C_n(M)$ or 
$\rm D_{n}(M)$. MS groups that are isomorphic to the improper rotation groups $\rm S_{2n} (M)$ with even $\rm n$ are exceptional: They can be written neither as a direct nor a 
semi-direct product of their permutation subgroup and one other subgroup of order two. Since these groups are cyclic, however, their permutation subgroups are necessarily cyclic 
as well; they are isomorphic to $\rm C_n (M)$. As we show in \autoref{sec:frozen-nuc-localised}, this is the only argument we need to prove that the frozen-nuclei approximation is 
physically incorrect.

The groups ${\rm C^{\perp}_s(M)}$ and ${\rm C_i(M)}$ in \autoref{Eq:Cnv-prod}, \autoref{Eq:Dnd-prod}, and \autoref{Eq:Sn-prod} are both of order two, and they contain 
besides the identity only one permutation-inversion. For the groups $\rm D_{nd}(M)$ and $\rm C_{nv}(M)$, one presentation of ${\rm C^{\perp}_s(M)}$ is
\be
\label{Eq:pres-c-perp-s}
{\rm C^{\perp}_s(M)}
=
\big<g_{\perp} \,\big|\, g^2_{\perp}={\rm E} \big>  \;,
\ee
where the generator $g_{\perp}$ is
\bse
\begin{align}
\label{Eq:pres-c-perp-s-gen-even}
g_{\perp}=&(2,{\rm n})(3,{\rm n}-1)...(\nicefrac{{\rm n}}{2},\nicefrac{{\rm n}}{2}+2)({\rm n}+1,2n)({\rm n}+2,2v-1)...(\nicefrac{3{\rm n}}{2},\nicefrac{3{\rm n}}{2}+1)^*\\
\intertext{if $\rm n$ is even, and} 
\label{Eq:pres-c-perp-s-gen-odd}
g_{\perp}=&(2,{\rm n})(3,{\rm n}-1)...(\nicefrac{{\rm n}+1}{2},\nicefrac{{\rm n}+3}{2})({\rm n}+2,2{\rm n})({\rm n}+3,2{\rm n}-1)...(\nicefrac{3{\rm n}+1}{2},\nicefrac{3{\rm n}+3}{2})^*
\end{align}
\ese
if $\rm n$ is odd. The presentation of the group ${\rm C_i(M)}$ with generator $g_{i}$ in \autoref{Eq:Sn-prod} is
\bse
\begin{align}
{\rm C_i(M)}
=&
\big<g_{i}=(1,{\rm n}+1)(2,{\rm n}+2)(3,{\rm n}+3) ...({\rm n},2{\rm n})^* \,\big|\, g^2_{i}={\rm E} \big>  
\end{align}
\ese
with $\rm n$ necessarily being odd;\autocite{Ezra.1982,Altmann.1994,Mirman.2007} see also \autoref{Fig:gruppe-abstr}.

Hence, as in case of the groups $\rm C_{nh}(M)$ and $\rm D_{nh}(M)$, the only groups that are relevant for our discussion are $\rm C_n(M)$ and $\rm D_{n}(M)$, and 
considering the permutation subgroup only, the results for molecules with MS groups $\rm S_n(M)$, $\rm C_{nv}(M)$ and $\rm D_{nd}(M)$ are the same as for molecules with 
$\rm C_{nh}(M)$, $\rm D_{nh}(M)$, $\rm C_{n}(M)$, and $\rm D_{n}(M)$ symmetry, respectively. While the operations of $\rm C_n(M)$, \textit{c.f.} \autoref{Eq:pres-Cn}, 
interconvert classical nuclear configurations that belong to different values of $\chi$, the operations of the group $\rm C^{\perp}(M)$, see \autoref{Eq:pres-C2-perp-1} and 
\autoref{Eq:pres-C2-perp-2}, transform different potential-dressed versions into one another.

As pointed out in \autoref{subsec:perm-sym-sym-top} of this Section, in terms of their alignment symmetric top molecules are very similar to linear molecules. Not only the 
potential, \autoref{Eq:pot-alg}, is identical for both type of rotors; see \autoref{Fig:pot-aus-st}. The structure of the permutation subgroup of their MS groups are identical, too. 
Setting ${\rm n}=1$ in \autoref{Eq:Dn-dir}, only the group $\rm C_2^{\perp}(M)$ remains as a symmetry group. Yet, this group is identical to the permutation subgroup of the 
$\rm D_{\infty,h}(M)$, which is the MS group of a linear molecule with two identical ends. Hence, the operation $(12)$ of this group changes the potential-dressed version of the 
linear molecule from $1$ to $2$ and \textit{vice versa}. The difference between symmetric tops and linear molecules is that for linear molecules only the two potential-dressed 
versions exists; there are no primed configurations for linear molecules. All other deliberations, however, are the same for both types of rigid rotors.

\thisfloatsetup{floatwidth=0.5\textwidth,capposition=beside}
\begin{figure}[tb!]
\centering{\includegraphics[width=0.475\textwidth]{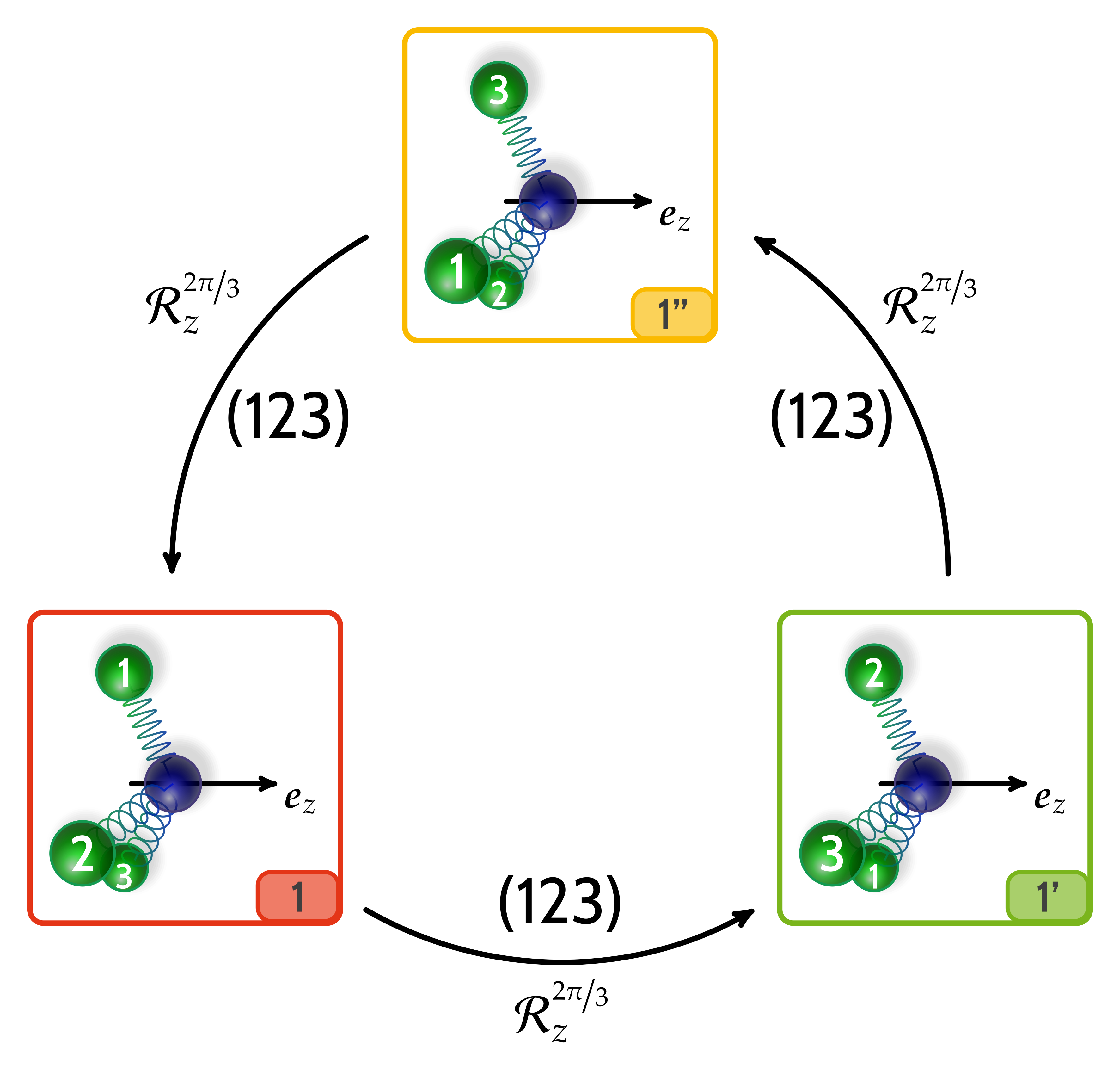}}


\caption{Three classical configurations of {NF$_3$} with MS group $\rm C_{3v}(M)$. They are interconverted by applying the permutation $(123)$ to the three fluorine nuclei.
Each configuration can be mimicked by a localised state.}
\label{Fig:nf3}
\end{figure}

\paragraph{Generalizations: polar symmetric tops -- and beyond}
A discussion of orienting polar symmetric tops leads to analogues results. For being able to orient molecules with an interaction of such as \autoref{Eq:ham-or}, the manipulated 
molecules must have a permanent dipole moment. As standard text books report,\autocite{Tinkham.2003,Heine.1993,McWeeny.2002,Bunker.1998} this is only possible if the 
molecule belongs to the MS groups $\rm C_n(M)$ or $\rm C_{nv}(M)$, see \autoref{Eq:pres-Cn}, \autoref{Eq:Cnv-prod}, \autoref{Eq:pres-c-perp-s}, 
\autoref{Eq:pres-c-perp-s-gen-even}, and \autoref{Eq:pres-c-perp-s-gen-odd} for their presentations. Further, only the set of permutations of identical nuclei---and thus, only 
$\rm G^{psms}$---remains a symmetry group of the system \textquote{molecule in electromagnetic field}, because the Hamiltonian $\mak H^{\rm or}$ is no longer invariant 
under permutation-inversions $\mal P^*$.\autocite{Watson.1975,Bunker.1998} The difference in energy for the aligned and anti-aligned molecule that \autoref{Fig:pot-or} shows 
reflects the loss in symmetry. Thus, there is only one potential-dressed version for polar symmetric tops; configuration $1$ and $2$ in \autoref{Fig:pot-or} are no longer 
energetically equivalent. 

Yet, there exist ${\rm n}-1$ primed configurations corresponding to the minimum of $\mak V^{\rm ad}$ in \autoref{Eq:ham-or-lin} that are transformed into each other by 
permuting (groups of) identical nuclei. Let us consider nitrogen trifluoride, NF$_3$, with MS group $\rm C_{3v}(M)$, which is an example for a symmetric top with permanent 
dipole moment; see also \autoref{Fig:nf3}. Even in its oriented configuration with $\theta=0$, there exist two more energetically configurations with
\bse
\bea
\chi_{1'} &=& \chi_1+\frac{2}{3}\pi\\
\chi_{1''} &=& \chi_1+\frac{4}{3}\pi\;.
\eea
\ese
Both also belong to the potential minimum at $\theta=0$, and they are interconverted by the permutations $(123)$ and $(132)$. In general, the number of permutationally 
equivalent configurations with $\theta=0$ is identical to the order of $\rm C_n$.

Since the Hamiltonian \autoref{Eq:ham-or-lin} is identical for symmetric and asymmetric tops, the same results apply to molecules with $\rm C_{2v}(M)$ and $\rm C_{2}(M)$ 
symmetry. Hence, for orienting asymmetric tops, we arrive at the same conclusion: The potential-dressed version associated with the minimum of the potential  
\autoref{Eq:ham-or} is not unique. Two configurations in $\chi$ exist, which both belong to the minimum at $\theta=0$. \Autoref{Fig:pot-or} illustrates this result for 
iodobenzene with MS group $\rm C_{2v}(M)$. 

Thus, for a dipole-type interaction such as \autoref{Eq:ham-or}, distinguishing the type of rotor is not necessary. The result is the same for all of them: As long as they posit a 
permutational symmetry with respect to the exchange of identical nuclei, the minimum of the potential \autoref{Eq:ham-or} cannot be identified with one classical nuclear 
configuration alone. There exists at least one more classical arrangement of the nuclei that also corresponds to this minimum and that results from the exchange (of groups) of 
identical nuclei.


\thisfloatsetup{floatwidth=0.575\textwidth,capposition=beside}
\begin{figure}[tb!]
\centering{\includegraphics[width=0.55\textwidth]{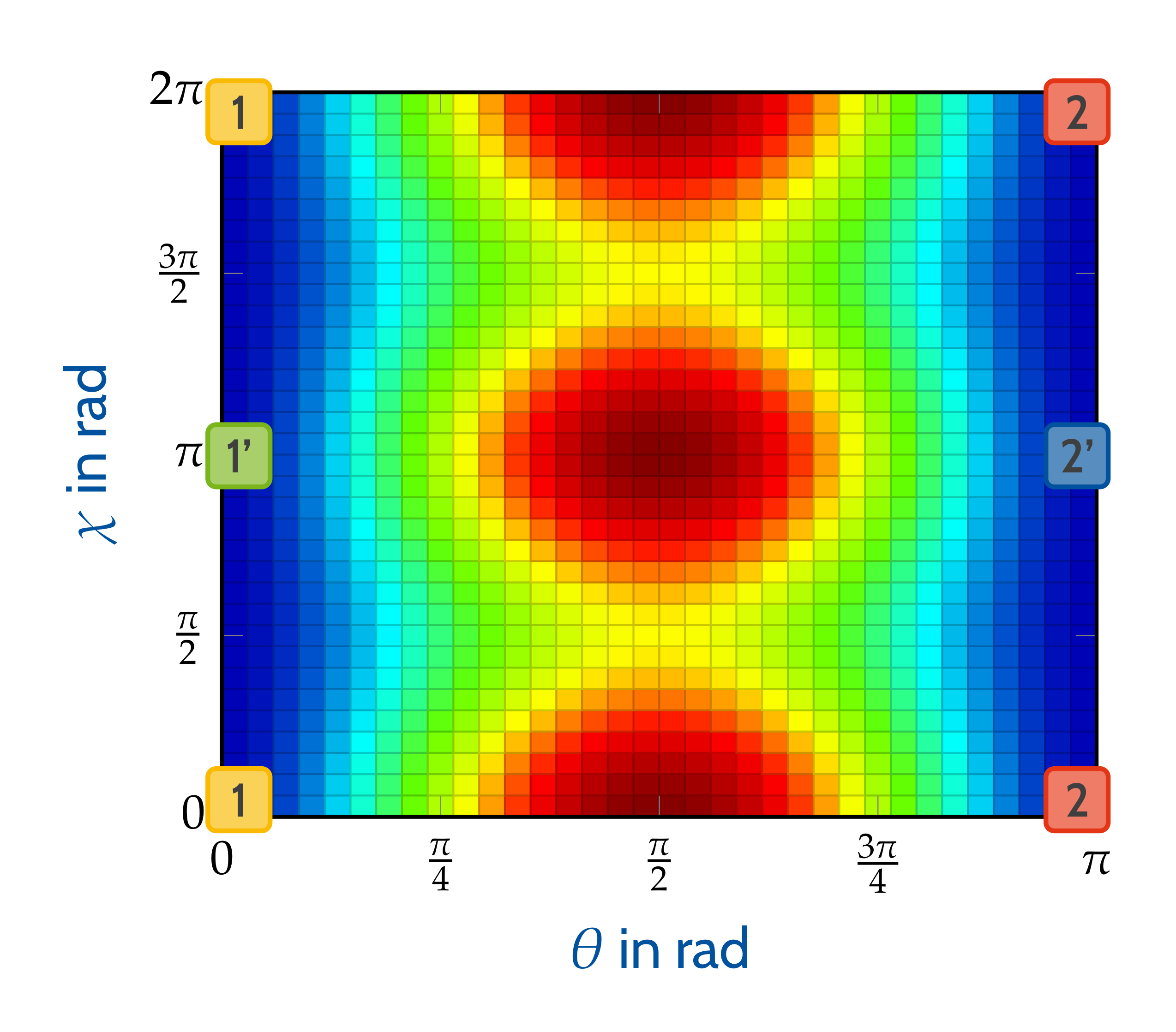}}


\caption{Contour plot of the adiabatic potential for the alignment of an asymmetric top. Depending on the symmetry, the four different minima at $(\theta=0,\chi=0)$, 
$(\theta=\pi,\chi=0)$, $(\theta=0,\chi=\pi)$, and $(\theta=\pi,\chi=\pi)$, belong to four different potential-dressed versions of the molecule; see \autoref{Fig:vers-C2F4} for the 
example of tetrafluoroethylene. If the barriers between the minima are sufficiently low, observable tunnelling among the different versions occurs.}
\label{Fig:pot-asym-top}
\end{figure}

\subsection{Confinement and symmetry: asymmetric tops}
\label{subsec:perm-sym-asym-top}
Asymmetric tops have, in general, a lower symmetry than symmetric tops. The asymmetric tops that we are interested in have MS groups $\rm D_{2h}(M)$, $\rm D_2(M)$, 
$\rm C_{2h}(M)$, $\rm C_{2v}(M)$, or $\rm C_2(M)$ only. This lower symmetry changes the potential for the alignment, as \autoref{Eq:ham-align-lin} and 
\autoref{Fig:pot-asym-top} show: If an asymmetric top is subject to a linearly-polarised laser pulse, the potential created by the field has four equivalent minima, each being 
distinguished by different classical configurations $(\theta,\chi)$. In particular, primed versions of the molecule are now separated by potential barriers, too.

\thisfloatsetup{floatwidth=0.4\textwidth,capposition=beside}
\begin{figure}[tb!]
\centering{\includegraphics[width=0.375\textwidth]{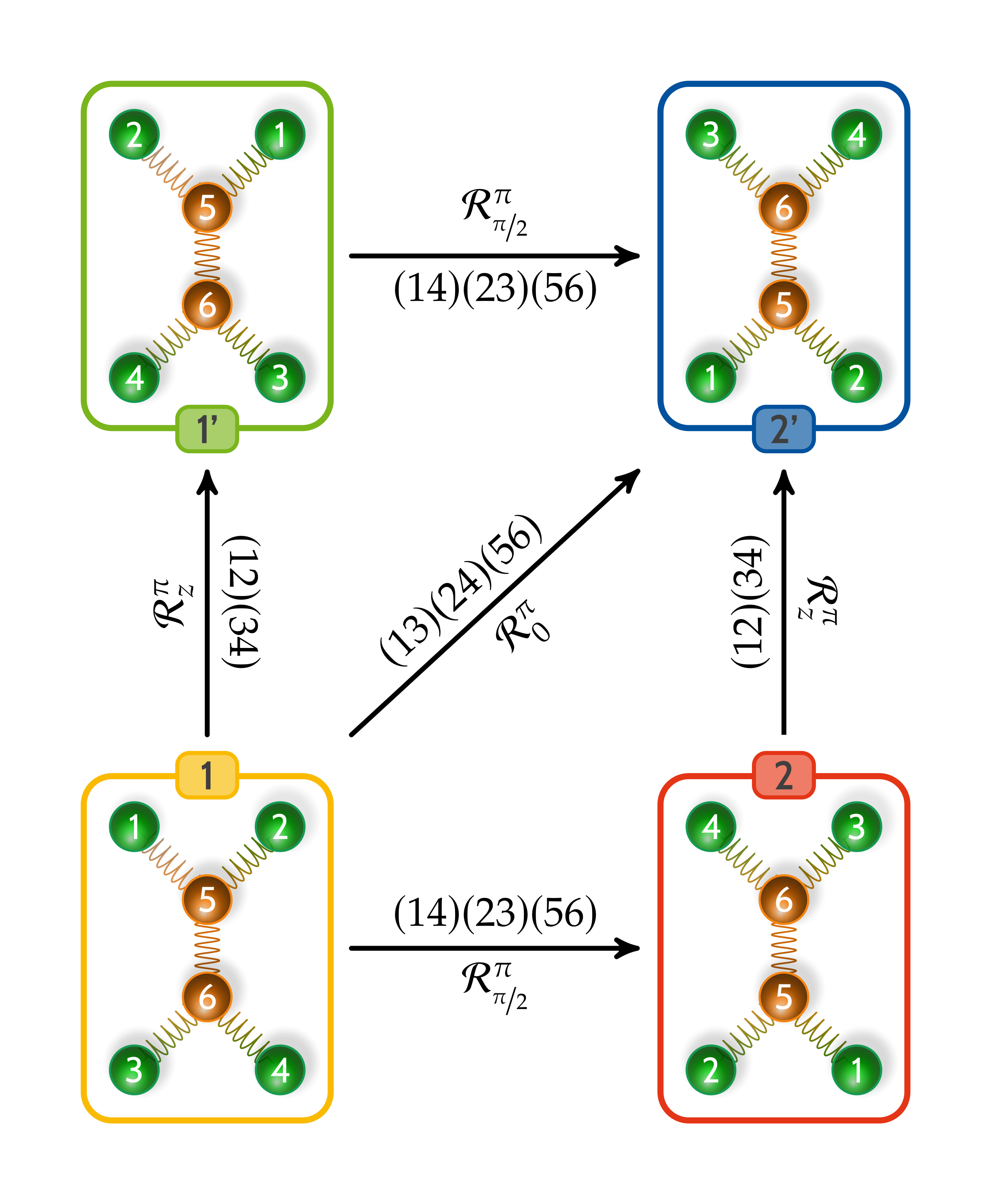}}


\caption{Illustration of the operations of the MS group $\rm D_{2}(M)$, applied to tetrafluoroethylene C$_2$F$_4$. The operations $(12)(34)$, $(14)(23)(56)$, and $(13)(24)(56)$ 
transform the potential-dressed version $1$ into the three potential-dressed versions $1'$, $2$, and $2'$, respectively. Hence, each of these permutations corresponds to a 
different equivalent rotation. Moreover, each potential-dressed version can be represented by a state being localised in the particular minimum.}
\label{Fig:vers-C2F4}
\end{figure}

\paragraph{Another illustrative example: tetrafluoroethylene}
This is maybe best seen by example. \Autoref{Fig:pot-asym-top} shows a contour plot of the potential for aligning an asymmetric top. For tetrafluoroethylene, the four 
equivalent minima correspond to four different potential-dressed versions, which are all separated by potential barriers created by the laser field. Depending on the barrier 
height, which is proportional to the field strength, tunnelling between these four versions might not be observed on the time-scale of the experiment.

Analysing the potential with the MS group shows that the four configurations are interchanged by the elements of the permutational subgroup of the MS group of 
tetrafluoroethylene. The MS group for tetrafluoroethylene, C$_2$F$_4$, in the electronic ground state is the same as for ethylene, which is $\rm D_{2h}(M)$. This group can be 
written as
\be
\label{Eq:D2h}
{\rm D_{2h}(M)}
=
{\rm G^{psms}} \otimes \left\{{\rm E},{\rm E}^*\right\}
\ee
with
\addtocounter{equation}{-1}
\bse
\be
\label{Eq:D2h-ps}
{\rm G^{psms}} \cong {\rm D_{2}(M)} = \left\{{\rm E},(12)(34),(13)(24)(56), (14)(23)(56)\right\}\;;
\ee
\ese
see \autoref{Fig:vers-C2F4} for the meaning of the permutations. Due to its structure, see \autoref{Eq:D2h}, the MS group of tetrafluoroethylene belongs to the MS groups of 
\textit{type-II}$^a$. Moreover, the group $\rm D_2(M)$ decomposes according to\autocite{Ezra.1982}
\be
{\rm D_{2}(M)}
=
{\rm C_2 (M)}
\otimes
{\rm C^{\perp}_2 (M)}
= 
\left\{{\rm E},(12)(34)\right\} \otimes \left\{(13)(24)(56), (14)(23)(56)\right\}\;,
\ee
where both subgroups, $\rm C_2(M)$ and $\rm C_2^{\perp}$, are cyclic. 

\Autoref{Fig:vers-C2F4} illustrates how the elements of $\rm D_2(M)$ transform the potential-dressed version $1$ into the other three potential-dressed versions $1'$, $2$, and 
$2'$, respectively. Hence, each permutation from $\rm D_2(M)$ belongs to a different equivalent rotation of the molecule. As in case of symmetric tops, applying the 
frozen-nuclei approximation is the same as choosing one of the four potential-dressed versions for describing the molecule. The difference between the two types of rotors is that 
between primed and non-primed versions a potential barrier exists, which might hinder the rotation about the molecule-fixed $\bm e_z$-axis.

\paragraph{Generalizations: polarizable asymmetric tops}
For molecules with MS group $\rm D_2(M)$, $\rm C_{2h}(M)$, $\rm C_{2v}(M)$ or $\rm C_2(M)$ similar conclusions hold. If the molecule belongs to $\rm D_2(M)$, the results are the 
same as for molecules with $\rm D_{2h}(M)$ symmetry. Because only permutations matter for our argument, and both $\rm D_{2h}(M)$ and $\rm D_2(M)$ have identical 
permutation subgroups, for molecules with $\rm D_2(M)$ symmetry, too, there exist four potential-dressed versions that are interconverted by the operations of 
$\rm G^{\rm psms}$. 

In case of molecules with MS group $\rm C_{2h}(M)$, $\rm C_{2v}(M)$, or $\rm C_2(M)$, the argument is less strict. All groups have the same abstract permutation subgroup, 
namely
\be
{\rm G^{\rm psms}}[{\rm C_{2h}(M)}]
\cong
{\rm G^{\rm psms}}[{\rm C_{2v}(M)}]
\cong
{\rm C_2(M)}
=\left\{{\rm E},(12)\right\}\;,
\ee
where
\be
{\rm C_{2h}(M)}
\cong
{\rm C_{2v}(M)}
=
{\rm C_2(M)}\otimes \left\{{\rm E},{\rm E^*}\right\}\;.
\ee
If the main principal axis is the $\bm e_z$-axis, the operation $(12)$ interchanges version $1$ and $1'$ as well as version $2$ and $2'$; see \autoref{Fig:pot-asym-top}. Going 
from minimum $1$ to minimum $2$, however, cannot be described in terms of permuting identical nuclei, although the minima $1$ and $2$ are systematically degenerate. The 
reason for that is the rigid-rotor approximation, according to which all asymmetric tops belong to the symmetry group $\rm D_2$.\autocite{Bunker.1998} Hence, even though 
the symmetry of the molecules is lower, the rigid rotor approximation imposes the symmetry group $\rm D_2$, and the minima on the potential that is created by the field are 
the same as for a molecule with $\rm D_{2}(M)$ or $\rm D_{2h}(M)$ symmetry. Yet, also for molecules with $\rm C_{2h}(M)$, $\rm C_{2v}(M)$, or $\rm C_2(M)$ applying the 
frozen-nuclei approximation is still equivalent to choosing one out of two potential-dressed versions of the molecule. 

\subsection{Why the frozen-nuclei argument works: localised states}
\label{subsec:loc-state-arg}
Studies on molecular control employing the two-step approach commonly use low-dimensional models, which assume that the molecule is perfectly oriented along one of its 
principal axes; see \autoref{Fig:two-step}.\autocite{Grohmann.2018c} The argument theoreticians use for eliminating most of the nuclear degrees of freedom in their models is 
that these motions can either be adiabatically separated from the manipulated degrees of freedom, or they can be considered as frozen on the time-scale of the studied process; 
see Refs. \citenum{Grohmann.2018c} and \citenum{Grohmann.2018b} for an elaborated discussion. In particular, they assume that it is legitimate to consider the molecule as being 
trapped in one potential minimum if only the strength of the external field is high enough. To calculate the time-evolution of the studied system, it is therefore sufficient to 
consider only the nuclear coordinates at one of the equivalent minima of the external potential.

A rationale for this argument gives the theory of MS groups, which is based on quantum tunnelling theory and the concept of structural degeneracy.\autocite{Bunker.1998} If the 
potential barriers between the ${\mak n}_{\rm ver}$ minima representing ${\mak n}_{\rm ver}$ structurally degenerate versions of the molecule is sufficiently large, the (low-lying) 
nuclear spectrum contains only sets of ${\mak n}_{\rm ver}$ quasi-degenerate energy eigenvalues. Hence, if the energy splittings within these sets are so small that they cannot be 
resolved in the experiment that is used to study the molecule, we can limit our theoretical considerations to the vicinity of only one of the ${\mak n}_{\rm ver}$ 
minima.\autocite{Bunker.2005} Then, all permutations and permutation-inversions interconverting the ${\mak n}_{\rm ver}$ versions are \textit{useless} for understanding the 
molecular spectrum and for characterizing electronic wave functions, as they lead to redundant selection rules. Consequently, it is sufficient to only consider the set of 
\textquote{feasible} nuclear permutations and permutation-inversions rather than the full nuclear permutation inversion group for analysing the molecular 
spectrum.\autocite{Bunker.1998,Bunker.2005}

Quantum dynamicists, however, not only care for energies, but they also study the time-evolution of molecular wave packets evolving on the potential energy surface. Thus, 
they have to modify the argument:\autocite{Grohmann.2012,Grohmann.2018b} Because the tunnelling times between different versions are much longer than the process we are 
interested in, we can form particular linear combinations of eigenstates such that they are localised in one of the ${\mak n}_{\rm ver}$ structurally degenerate minima. Hence, it is 
sufficient to only consider wave functions that are localised in the vicinity of one minimum, and when describing the time-evolution of molecular states, we can then use a local 
Schrödinger equation that contains a truncated potential only.

Both arguments are, at least implicitly, the starting point of quantum chemical and spectroscopic studies on the one hand, and the description of the quantum and classical 
dynamics of molecules on the other hand.\autocite{Grohmann.2012} But they also apply if molecules are subject to a strong external potential: If the potential barriers created by 
the external interaction are sufficiently high, energy splittings due to tunnelling between different potential-dressed versions of the molecule cannot be resolved. Hence, it is 
legitimate to consider only one minimum of the potential when calculating the spectrum of the molecule in the potential; see \autoref{Fig:pot-aus-st} and 
\autoref{Fig:pot-asym-top} for the aligning potentials of a symmetric and asymmetric top, respectively. Likewise, we can describe the quantum dynamics of the system in terms of 
states that are localised in one minimum, being linear combinations of wave functions belonging to quasi-degenerate energies.

Thus, if this assumption is right, it seems to be only a small approximation to replace the localised nuclear wave packet by a single \textquote{frozen} nuclear configuration. If the 
potential barriers are insuperable, and the process we want to describe is fast compared to the hindered rotational motion of the nuclei, this configuration will not change much on 
the relevant time-scale. Hence, the frozen-nuclei approximation appears to be reasonable as long as the rotational motion of the nuclei is slow compared to the time-scale of the 
studied phenomenon. 

This reasoning, however, is wrong. To begin with, our discussion in this Section shows that freezing the nuclei was only valid if \textit{all} classical configurations in the rotational 
coordinates are separated by insuperable barriers. Thus, for symmetric tops, aligning or orienting the molecule would be not sufficient to apply the frozen-nuclei approximation. All 
primed configurations of symmetric top molecules are not separated by barriers at all, see \autoref{Fig:pot-aus-st} for the example of benzene, and there are no reasonable grounds 
for assuming that choosing one primed configuration is legitimate; see in particular the discussion in \autoref{subsec:perm-sym-sym-top} of this Section.

Yet, there are more problems with this rationale, even if it was possible to separate each classical configuration by insuperable barriers created by some sort of external interaction. 
In particular, the localised rotational states on which low-dimensional models are relying cannot be realised. As we show in \autoref{subsec:perm-sym-sym-top} and  
\autoref{subsec:perm-sym-asym-top} of this Section, in many cases going from one structurally degenerate minimum of the external potential to another is equivalent to permuting 
identical nuclei. Under these circumstances, however, localised states are unphysical. Because they are linear combinations of eigenstates belonging to different nuclear spin 
isomers, they cannot be formed by an external interaction on reasonable time-scales, as we show in the following.

\section{Frozen nuclei and localised states}
\label{sec:frozen-nuc-localised}
Recently, we have extended the list of shortcomings molecular dynamics simulations entail. For internal molecular motions that can be described in terms of permutations of 
identical nuclei, such as torsions and pseudo-rotations, molecular dynamics simulations are inherently wrong. They (implicitly) presuppose that a localised wave packet is an 
adequate description of the quantum motion the simulations are supposed to mimic. For cyclic contortional motions, however, such states are 
unphysical.\autocite{Grohmann.2018b} 

In the following, we show that for the {frozen-nuclei approximation} identical conclusions hold. It implicitly assumes that localised states are physical representations of 
quantum systems, but for molecules with identical nuclei such states represent superpositions of (eigen-)states belonging to different nuclear spin isomers. Thus, if the nuclear spin 
isomer hypotheses is legitimate, localised states must be discarded as physical descriptions of the molecule---and hence, the {frozen-nuclei approximation}.   

It turns out that these similarities are not a coincidence: both concepts, molecular dynamics and the frozen-nuclei approximation, are based on a theory reduction rather than an 
approximation. By limiting the motion of nuclei to trajectories with or without zero momentum, they treat the nuclei as classical particles, and they therefore cannot account for 
effects that are inherently quantum mechanical.

\subsection{A definition of nuclear spin isomers of molecules}
According to the spin-statistics theorem, any molecular wave function $\Phi^{\rm mol}$ must be either symmetric ($\zeta^{\rm mol}=+1$) or anti-symmetric 
($\zeta^{\rm mol}=-1$) if two or more identical nuclei are permuted, where the character $\zeta^{\rm mol}$ depends on the spin of the permuted nuclei and the order of the 
permutation. Consequently, molecular wave functions must transform according to an irreducible, one-dimensional, and real molecular representation $\Gamma^{\rm mol}$ in 
the complete nuclear permutation group, $\rm G^{\rm cnp}$.\autocite{Bunker.1998} Since not all permutations cause observable tunnelling splittings, however, classifying 
molecular wave functions according to the permutation subgroup of the MS group, $\rm G^{\rm psms}$, is sufficient.\autocite{LonguetHiggins.1963,Watson.1965,Bunker.1998}

For most molecular systems known so far, writing any molecular eigenstate $\Phi^{\rm mol}$ as
\be
\label{Eq:nuc-spin-hyp}
\Phi^{\rm mol} = \Phi^{\rm rcve}\cdotp\Phi^{\rm nu.sp}\;,
\ee
is an excellent approximation.\autocite{Levitt.2008,Bunker.1998,Bunker.2009,Quack.2011,Chapovsky.1999} In \autoref{Eq:nuc-spin-hyp}, $\Phi^{\rm rcve}$ denote the 
eigenfunctions for the rotational-contorsional-vibrational-electronic motions of the molecule, and $\Phi^{\rm nu.sp}$ are the wave functions describing its nuclear spins. 

The nuclear spin isomer hypotheses combines the assumption \autoref{Eq:nuc-spin-hyp} and the spin-statistics theorem: If nuclear spin conversion effects are negligible, molecules 
exist in form of nuclear spin isomers,\autocite{Chapovsky.1999,Bunker.2009} which are determined by the equation\autocite{Bunker.1998,Haase.22.11.2007}
\be
\label{Eq:def-nsi-full}
\Gamma^{\rm mol} \subseteq \Gamma^{\rm rcve}\otimes \Gamma^{\rm nu.sp}\;.
\ee
Consequently, different nuclear spin isomers of a molecule are characterized by unique combinations of the irreducible representations $\Gamma^{\rm rcve}$ and 
$\Gamma^{\rm nu.sp}$ in the permutation subgroup of the MS group.\autocite{Grohmann.2018b,Haase.22.11.2007}

In experiments on rotational control, it is furthermore reasonable to assume that \textit{(i)} the rotations separate adiabatically from the remaining motions of the 
molecule;\autocite{Friedrich.1991,Seideman.2005,Grohmann.2018b} and \textit{(ii)} any external laser field does not manipulate the quantum states of the other 
motions.\autocite{Friedrich.1991,Seideman.2005,Grohmann.2018b} Under these conditions, we can formally average over the electronic and vibrational coordinates and replace 
$\Phi^{\rm mol}$ by the model wave functions\autocite{Grohmann.2018b}
\be
\label{Eq:phi-mod}
\Phi^{\rm mod} = \Phi^{\rm rot}(\theta,\phi,\chi)\cdotp \Phi^{\rm nu.sp}(\bm\Sigma)\;.
\ee
In \autoref{Eq:phi-mod}, $\Phi^{\rm rot}$ are the rotational (eigen-)functions, which are dependent on the Euler angles $\theta,\phi,\chi$, and $\Phi^{\rm nu.sp}$ denote the 
nuclear spin eigenstates as a function of the collective nuclear spin variable $\bm\Sigma$. If the assumptions \textit{(i)} and \textit{(ii)} are both acceptable, the wave functions 
describing the rotational motion are, or can be expanded into the eigenfunctions of a symmetric top.\autocite{Bunker.1998} Both the rotational and the nuclear spin functions in 
\autoref{Eq:phi-mod} individually form a basis for the irreducible representations of the permutation subgroup of the MS group.\autocite{Bunker.1998,Bunker.2005} 

Using the adiabatic ansatz of separated molecular time-scales and focussing on the rotational motion of the nuclei, \autoref{Eq:def-nsi-full} reduces to\autocite{Grohmann.2018b}
\be
\label{Eq:def-nsi}
\Gamma^{\rm mod} \subseteq \Gamma^{\rm rot} \otimes \Gamma^{\rm nu.sp}\;.
\ee
In \autoref{Eq:def-nsi}, $\Gamma^{\rm mod}$, $\Gamma^{\rm rot}$, and $\Gamma^{\rm nu.sp}$ are spanned by the (symmetry-adapted) wave functions $\Phi^{\rm mod}$, 
$\Phi^{\rm rot}$, and $\Phi^{\rm nu.sp}$, respectively. As in the general case, \autoref{Eq:def-nsi-full}, the irreducible representation $\Gamma^{\rm mod}$ in \autoref{Eq:def-nsi} 
is fixed and dictated by the spin-statistics theorem. Hence, the nuclear spin isomers of the rotating nuclear frame are unambiguously defined by a combination of rotational and 
nuclear spin symmetries, which we denote as $\Gamma^{\rm rot}[\Gamma^{\rm nu.sp}]$ hereafter.

\subsection{On localised rotational states in external potentials}
To generalize our earlier results for cyclic, one-dimensional, contortional motions of non-rigid molecules\autocite{Grohmann.2018b} to the rotations of rigid molecules, we 
distinguish two cases: \textit{(i)} the permutation subgroup of the MS group of the rigid molecule is cyclic; and \textit{(ii)}  the permutation subgroup of the MS group can be 
decomposed into cyclic groups. In both cases the results are the same: (partially) localised states are superpositions of states belonging to different irreducible representations of 
cyclic subgroups of the permutation subgroup of the MS group.

\paragraph{The permutation subgroup of the MS group is cyclic}
In the simplest case, $\rm G^{\rm psms}$ is identical to the group 
\be
\label{Eq:Cn}
{\rm C_n}({\rm M}) = \big<g=(12 \hdots {\rm n}) \,\big|\, g^{\rm n} = {\rm E} \big> \;,
\ee
where $g$ denotes the generator of ${\rm C_n}({\rm M})$; $\rm E$ is the identity; and $\rm n$ gives the number of (groups of) identical nuclei that are permuted as the molecule 
rotates. Because ${\rm C_n}({\rm M})$ is cyclic, the irreducible representations $\Gamma$ of this group are all one-dimensional.\autocite{McWeeny.2002} As 
\autoref{sec:sym-2step}, \autoref{subsec:perm-sym-sym-top} and \autoref{subsec:perm-sym-asym-top} show, $\rm G^{\rm psms}$ has exactly this structure if the rigid molecule 
belongs to the MS groups $\rm C_{nh}(M)$, $\rm C_{nv}(M)$, $\rm C_n (M)$, or $\rm S_{2n}(M)$. 

Let us consider NF$_3$ as an example. In the electronic ground state, this molecule belongs to the MS group $\rm C_{3v}(M)$ with permutation subgroup 
$\rm C_3(M)$.\autocite{Bunker.2005} Hence, ${\rm n}=3$ in \autoref{Eq:Cn} and one generator of this group is $(123)$; see \autoref{Fig:nf3} for an illustration. In case the molecule 
is oriented, three localised states exist; they belong to the same field-dressed version, but to different classical configurations $(\theta_1,\chi_1)$, 
$(\theta_{1'}=\theta_1,\chi_{1'}=\chi_{1}+\nicefrac{2\pi}{3})$, and $(\theta_{1''}=\theta_1,\chi_{1''}=\chi_{1}+\nicefrac{4\pi}{3})$. Applying the frozen-nuclei approximation is 
equivalent to choosing one of the three. The permutation $(123)$ interconverts the three configurations; see \autoref{Fig:nf3}. Thus, applied to a state $\Phi^{\rm loc}_1$ 
localised at $(\theta_1,\chi_1)$, the operation $(123)$ generates a state $\Phi^{\rm loc}_2$ that is localised at $\chi_{1}+\nicefrac{2\pi}{3}$,  while $(123)(123)$ gives a state 
$\Phi^{\rm loc}_3$ that is localised at $\chi_{1}+\nicefrac{4\pi}{3}$.

The localised states $\Phi^{\rm loc}_j$, $j=1,2,3$, are linear combinations of symmetry-adapted rotational states, \textit{i.e.} states that transform irreducible in the group 
$\rm C_n(M)$. This follows from the fact that any field-dressed state created by the field can always be expanded in terms of symmetry-adapted symmetric top 
eigenfunctions\autocite{McWeeny.2002} 
\be
\label{Eq:eig-st}
\Phi^{\rm fd}_{{\mak n}_{\rm pen},\Gamma} (\theta,\phi,\chi)
= 
\sum_{J,k,m} c_{J,k,m;\Gamma,{\mak n}_{\rm pen}} {\mak d}^{J}_{-m,-k} (\theta)\cdotp \exp\left({\rm i}m\phi\right)\cdotp\exp\left({\rm i}k\chi\right)\;.
\ee
In \autoref{Eq:eig-st}, $J,k,m$ denote the quantum numbers of a symmetric top; $c_{J,k,m;\Gamma,{\mak n}_{\rm pen}}$ are the expansion coefficients of the 
$\mak n_{\rm pen}$-th pendular state with symmetry $\Gamma$; and ${\mak d}^{J}_{-m,-k}$ represents Wigner's small d-matrix.\autocite{Zare.1988} Using the projection operator 
technique,\autocite{McWeeny.2002} and taking into account the transformation properties of the rotational wave functions from \autoref{Eq:eig-st} in the MS 
group,\autocite{Haase.22.11.2007,Bunker.1998} we obtain\autocite{Grohmann.2018b}
\be
\label{Eq:loc-NF3}
\Phi^{\rm loc}_j
=
\frac{1}{\sqrt{3}}
\left(\Phi^{\rm rot}_{{\rm A}} + \epsilon_j\cdotp\Phi^{\rm rot}_{{\rm E_1}} + \epsilon_j^2\cdotp\Phi^{\rm rot}_{{\rm E_2}}\right)\quad j=1,2,3.
\ee
In \autoref{Eq:loc-NF3}, $\epsilon_j$ is a complex number with $|\epsilon_j|^2=1$, and $\Phi^{\rm rot}_{{\rm A}}$, $\Phi^{\rm rot}_{{\rm E_1}}$ and $\Phi^{\rm rot}_{{\rm E_2}}$ are 
rotational wave functions belonging to the irreducible representations $\rm A$, $\rm E_1$, and $\rm E_2$ in $\rm C_3(M)$, respectively, formed by the three localised 
states $\Phi^{\rm loc}_j$.\autocite{Grohmann.2018b} Hence, each of the three localised states is a linear combination of symmetry-adapted states and \textit{vice versa}; see Ref. 
\citenum{Grohmann.2018b} for a derivation.

We stress that the symmetry-adapted states in \autoref{Eq:loc-NF3} are not necessarily eigenfunctions; they might be time-dependent symmetry-adapted wave packets instead. 
For our argument is only important, however, that no matter what shape the localised state has, it is always a linear combination of three symmetry-adapted states.  The only 
condition the three symmetry-adapted states have to fulfil for $\Phi^{\rm loc}_j$ to be localised is
\be
|\Phi^{\rm rot}_{{\rm A}}|^2 
\buildrel ! \over = 
|\Phi^{\rm rot}_{{\rm E_1}}|^2 
\buildrel ! \over =
|\Phi^{\rm rot}_{{\rm E_2}}|^2\;.
\ee
Hence, all symmetry-adapted states contribute with the same weight to any of the three localised states.

These arguments are valid for any arbitrary cyclic $\rm G^{psms}$: For a system with the symmetry group ${\rm C_n}({\rm M})$, each localised rotational state is a superposition of 
symmetry-adapted rotational functions. Only if one symmetry-adapted state for each irreducible representation contributes with the same weight to $\Phi^{\rm loc}$ in the sense 
of \autoref{Eq:loc-NF3}, the state is localised exactly at one classical configuration.\autocite{Grohmann.2018b}

\paragraph{The permutation subgroup of the MS group decomposes into cyclic groups}
In general, the group $\rm G^{psms}$ is not cyclic. Rigid molecules whose permutation subgroup is not cyclic itself have to belong to the MS groups $\rm D_n(M)$, $\rm D_{nh}(M)$, 
or $\rm D_{nd}(M)$, as our analysis in \autoref{sec:sym-2step} shows. Yet, in all of these cases, the permutation subgroup $\rm G^{psms}$ is isomorphic to $\rm D_n(M)$, which can 
always be decomposed into cyclic subgroups, see \autoref{sec:sym-2step}, \autoref{subsec:perm-sym-sym-top} and \autoref{subsec:perm-sym-asym-top}. Formally, we can write
\bse
\bea
\label{Eq:msps-non-cyc-2}
{\rm D_2 (M)} &=& {\rm C_2(M)} \otimes  {\rm C^{\perp}_2(M)} \\
\label{Eq:msps-non-cyc-n}
{\rm D_n (M)} &=& {\rm C_n(M)} \rtimes  {\rm C^{\perp}_2(M)} \quad\text{if}\;  {\rm n} >2\;,
\eea
\ese
where the operations of $\rm C_n(M)$ can be mapped on the equivalent rotations $\mal R^{\beta}_z$, only changing $\chi$, and the operations of $ {\rm C^{\perp}_2(M)}$ can be 
mapped on the equivalent rotation $\mal R^{\pi}_{\alpha}$, only changing $\phi$ and $\theta$; see \autoref{Eq:R-eq-alph} and  \autoref{Eq:R-eq-beta}, respectively, and 
\autoref{subsec:oper-ms} of the Appendix. 

Due to the decomposition of the permutation subgroup, see \autoref{Eq:msps-non-cyc-2} and \autoref{Eq:msps-non-cyc-n}, and the form of the rotational basis functions, 
\textit{c.f.} \autoref{Eq:eig-st}, we can generalize the results from the last paragraph. Because the rotational basis functions are a product of functions being dependent 
on only one rotational coordinate, we can analyse the symmetry and the localisation of the rotational wave functions separately in each coordinate and each cyclic subgroup of 
$\rm G^{psms}$. 

A simple example is aligned C$_2$F$_4$ with MS group $\rm D_{2h}(M)$.\autocite{Grohmann.2011} Here, four localised classical configurations---$1$, $1'$, $2$, and $2'$---are 
separated by the potential created by a linearly-polarized aligning field; see \autoref{Fig:vers-C2F4} and the discussion in \autoref{sec:sym-2step}, 
\autoref{subsec:perm-sym-asym-top}. The permutation subgroup $\rm G^{psms}$ is $\rm D_2(M)$; see \autoref{Eq:D2h} and \autoref{Eq:D2h-ps}. As it is clear from 
\autoref{Fig:pot-asym-top} and \autoref{Fig:vers-C2F4}, the operations of 
\bse
\be
\label{Eq:C2F4-psms-1}
{\rm C_2(M)}= \left\{ {\rm E},(12)(34) \right\}
\ee
interconvert field-dressed version $1$ and $1'$, only changing $\chi$ to $\chi+\pi$. Conversely, the operations of 
\be
\label{Eq:C2F4-psms-2}
{\rm C^{\perp}_2(M)}=\left\{{\rm E},(14)(23)(56) \right\}
\ee
\ese
change field-dressed version $1$ to field-dressed version $2$ and $\theta$ to $\pi - \theta$, respectively. 

For the localisation in $\chi$, the result is the same as for NF$_3$: Any rotational wave function that is localised in $\chi$ is a superposition of symmetry-adapted 
rotational functions, classified according to the subgroup $\rm C_2(M)$ of the permutational subgroup with irreducible representations $\rm A_{\chi}$ and 
$\rm B_{\chi}$.\autocite{Bunker.1998,Altmann.1994} Consider two states $\Phi^{\rm loc}_1$ and $\Phi^{\rm loc}_{1'}$ that are localised at minimum $1$ and $1'$ of the potential 
in \autoref{Fig:pot-asym-top}, respectively. For these states hold 
\bse
\begin{align}
\begin{split}
(12)(34)\, \Phi^{\rm loc}_1(\theta,\phi,\chi)
&= \Phi^{\rm loc}_1(\theta,\phi,\chi+\pi)\\ 
&= \Phi^{\rm loc}_{1'} (\theta,\phi,\chi)
\end{split}
\\
\begin{split}
(12)(34)\, \Phi^{\rm loc}_{1'}(\theta,\phi,\chi)
&= \Phi^{\rm loc}_{1'}(\theta,\phi,\chi+\pi)\\
&= \Phi^{\rm loc}_{1} (\theta,\phi,\chi)\;.
\end{split}
\end{align}
\ese
If we apply the projection operators of $\rm C_2(M)$ to the localised states,\autocite{Grohmann.2018b} we obtain the symmetry-adapted functions
\bse
\bea
\label{Eq:sym-ad-wf-chi-1}
\Phi^{\rm A_{\chi}}
&=&
\frac{1}{\sqrt{2}}
\left(\Phi^{\rm loc}_{1} + \Phi^{\rm loc}_{1'} \right)\\
\label{Eq:sym-ad-wf-chi-2}
\Phi^{\rm B_{\chi}}
&=&
\frac{1}{\sqrt{2}}
\left(\Phi^{\rm loc}_{1} - \Phi^{\rm loc}_{1'} \right) \;.
\eea
\ese
Thus,
\bse
\bea
\label{Eq:loc-wf-chi-1}
\Phi^{\rm loc}_{1}
&=&
\frac{1}{\sqrt{2}}
\left(\Phi^{\rm A_{\chi}} + \Phi^{\rm B_{\chi}} \right)\\
\label{Eq:loc-wf-chi-2}
\Phi^{\rm loc}_{1'}
&=&
\frac{1}{\sqrt{2}}
\left(\Phi^{\rm A_{\chi}} - \Phi^{\rm B_{\chi}} \right) \;.
\eea
\ese
Consequently, the two localised functions $\Phi^{\rm loc}_1$ and $\Phi^{\rm loc}_{1'}$ are linear combinations of rotational wave functions that transform irreducible in 
$\rm C_2(M)$.

Analogously, we can consider two states $\Phi^{\rm loc}_1$ and $\Phi^{\rm loc}_{2}$ that are localised in minimum $1$ and $2$ of the potential in \autoref{Fig:pot-asym-top}, 
respectively, and classify them according the irreducible representations $\rm A_{\theta}$ and ${\rm B}_{\theta}$ of the group 
${\rm C^{\perp}_2(M)}$.\autocite{Bunker.1998,Altmann.1994} Here,
\bse
\begin{align}
\begin{split}
(14)(23)(56)\, \Phi^{\rm loc}_1(\theta,\phi,\chi)
&= \Phi^{\rm loc}_1(\pi-\theta,\phi+\pi,\chi)\\ 
&= \Phi^{\rm loc}_{2} (\theta,\phi,\chi)
\end{split}
\\
\begin{split}
(14)(23)(56)\, \Phi^{\rm loc}_{2}(\theta,\phi,\chi)
&= \Phi^{\rm loc}_{2}(\pi-\theta,\phi+\pi,\chi)\\
&= \Phi^{\rm loc}_{1} (\theta,\phi,\chi)\;.
\end{split}
\end{align}
\ese
Applying the projections operators for $\rm A_{\theta}$ and ${\rm B}_{\theta}$ to $\Phi^{\rm loc}_1$, we obtain
\bse
\bea
\label{Eq:sym-ad-wf-theta-1}
\Phi^{\rm A_{\theta}}
&=&
\frac{1}{\sqrt{2}}
\left(\Phi^{\rm loc}_{1} + \Phi^{\rm loc}_{2} \right)\\
\label{Eq:sym-ad-wf-theta-2}
\Phi^{\rm B_{\theta}}
&=&
\frac{1}{\sqrt{2}}
\left(\Phi^{\rm loc}_{1} - \Phi^{\rm loc}_{2} \right)
\eea
\ese
and thus,
\bse
\bea
\label{Eq:loc-wf-theta-1}
\Phi^{\rm loc}_{1}
&=&
\frac{1}{\sqrt{2}}
\left(\Phi^{\rm A_{\theta}} + \Phi^{\rm B_{\theta}} \right)\\
\label{Eq:loc-wf-theta-2}
\Phi^{\rm loc}_{2}
&=&
\frac{1}{\sqrt{2}}
\left(\Phi^{\rm A_{\theta}} - \Phi^{\rm B_{\theta}} \right)\;.
\eea
\ese
Hence, also rotational states being localised in $\theta$ are superpositions of symmetry-adapted states, if classified according to the irreducible representations of 
${\rm C^{\perp}_2(M)}$.

What we have shown here for C$_2$F$_4$ is valid for any rigid molecule whose permutation subgroup, $\rm G^{psms}$, decomposes into cyclic subgroups. In general, the ensemble 
of localised states that are interconverted by the operations of one cyclic subgroup is a basis for the regular representation of this group;\autocite{Grohmann.2018b} they generate 
a representation that contains each irreducible representation of the group once and with equal weight. Consequently, it is always possible to construct symmetry-adapted 
rotational states out of localised rotational states, and each of these symmetry-adapted rotational states is a superposition of all localised rotational states that are interconverted 
by the operations of the respective cyclic subgroup of $\rm G^{psms}$.\autocite{Grohmann.2018b} 

Moreover, the functions that transform irreducible in the cyclic subgroups of $\rm G^{psms}$, such as \autoref{Eq:sym-ad-wf-chi-1}, \autoref{Eq:sym-ad-wf-chi-2}, 
\autoref{Eq:sym-ad-wf-theta-1}, and \autoref{Eq:sym-ad-wf-theta-2}, do not have to transform irreducible in the permutation subgroups or the full MS group. To show that localised 
states are unphysical representations of molecules with feasible permutations of identical nuclei, it suffices to proof that they are linear combinations of functions that transform 
irreducible in the subgroups $\rm C_n(M)$ and ${\rm C^{\perp}_2(M)}$, as the following discussion shows.

\subsection{The impossibility of localised states revisited}
So far we have only shown that localised rotational states are superposition states of different symmetries, if they are classified according to cyclic subgroups of the permutation 
subgroup of the MS group of the molecule. To make plausible why they represent unphysical states of the molecule, we need to discuss the following three aspects:

\ben
[leftmargin=20pt,
labelwidth=15pt,
labelsep=5pt,
rightmargin=0pt,
topsep=0\baselineskip,
itemsep=0.5\baselineskip,
align=right,
label*=\protect\arabic{enumi}),
format={\itshape\color{DFG-blau}}
]
\item
If the nuclei that are permuted by the operations of a cyclic subgroup of $\rm G^{psms}$ carry zero spin, only one nuclear spin isomer exists in this subgroup. If the nuclei carry 
non-zero spin, every possible nuclear spin isomer exists, \textit{i.e.} every symmetry-allowed combination of nuclear spin states and spatial states is realised.

\item
If the nuclei that are permuted by the operations of the cyclic subgroup carry zero spin, localised states are symmetry-forbidden. If the nuclei carry non-zero spin, spatially confined 
nuclear wave packets are superpositions of eigenstates belonging to different nuclear spin isomers.

\item
If couplings and correlations between nuclear spins and rotational motions are negligible, and the existence of stable nuclear spin isomers is a reasonable assumption, localised 
states are unphysical. They cannot be created by external interactions on the time-scale of typical experiments on molecular control if the nuclei are treated as quantum objects.
\een

In the following, we focus on the physical implications each of these statements have for the localisation of nuclei in space; the relevant proofs we have presented 
elsewhere.\autocite{Grohmann.2018b}

\paragraph{On the symmetry of nuclear spin states}
Let us consider the case of $\rm G^{psms}$ being cyclic first. A molecule for which the operations of a cyclic $\rm G^{psms}$ only permutes nuclei with zero spin, the nuclear spin 
functions always transform according to the totally symmetric representation, $\Gamma_{\rm ts}$, in $\rm G^{psms}$.\autocite{Grohmann.2018b} This argument is valid even if the 
molecule contains nuclei with non-zero spin, but which are not permuted by the operations of $\rm G^{psms}$. 

If, however, the permuted nuclei carry non-zero spin, for each irreducible representation of $\rm G^{psms}$, we find at least one nuclear spin function that forms a basis for this 
representation. In other words, the representation that is generated by the set of all nuclear spin functions, $\Gamma^{\rm nu.sp}$, always contains the regular representation of 
$\rm G^{psms}$.\autocite{Grohmann.2018b}

Although the arguments in Ref. \citenum{Grohmann.2018b} are exemplified for single nuclei, they equally apply to molecules for which the permutations of $\rm G^{psms}$ refer 
to rigid fragments of the molecule, like for iodobenzene with $\rm G^{psms}\cong C_{2}(M)$; see \autoref{Fig:pot-or}. Then, however, the representation of $\Gamma^{\rm nu.sp}$ 
is identical to $\Gamma_{\rm ts}$ only if \textit{all} nuclei that are permuted by the operations of $\rm G^{psms}$ carry non-zero spin. Otherwise, $\Gamma^{\rm nu.sp}$ always 
contains the regular representation of $\rm G^{psms}$.

If $\rm G^{psms}$ is not cyclic, for rigid molecules, it is still possible to decompose $\rm G^{psms}$ into cyclic subgroups; see \autoref{Eq:msps-non-cyc-2} and 
\autoref{Eq:msps-non-cyc-n} as well as the discussion in \autoref{sec:sym-2step}, \autoref{subsec:perm-sym-sym-top} and \autoref{subsec:perm-sym-asym-top}. Therefore, the 
arguments of the preceding paragraph also apply to the molecules of this type, because we can analyse their nuclear spin states separately in each cyclic subgroup of 
$\rm G^{psms}$. 

An illustrative example is again C$_2$F$_4$. We consider natural isotopes, $\mal I({\rm ^{12}C})=0$ and $\mal I({\rm ^{19}F})=\nicefrac{1}{2}$, and specify every non-symmetrised 
nuclear spin state by the quadruplet ($m_{\mal I_1}$,$m_{\mal I_2}$,$m_{\mal I_3}$,$m_{\mal I_4}$) with $m_{\mal I_i}=\pm \nicefrac{1}{2}$; see \autoref{Fig:vers-C2F4} for the 
labelling of identical nuclei. These sixteen nuclear spin states form a basis for the representations of the groups $\rm C_2(M)$ and ${\rm C^{\perp}_2(M)}$ from 
\autoref{Eq:C2F4-psms-1} and \autoref{Eq:C2F4-psms-2}, respectively.\autocite{Bunker.1998,Grohmann.2018b} 

Consider, for example, the nuclear spin states $(-\nicefrac{1}{2},\nicefrac{1}{2},\nicefrac{1}{2},-\nicefrac{1}{2})$ and $(\nicefrac{1}{2},-\nicefrac{1}{2},-\nicefrac{1}{2},\nicefrac{1}{2})$. 
Since
\bse
\bea
(-\nicefrac{1}{2},\nicefrac{1}{2},\nicefrac{1}{2},-\nicefrac{1}{2}) &\xrightarrow{(12)(34)}& (\nicefrac{1}{2},-\nicefrac{1}{2},-\nicefrac{1}{2},\nicefrac{1}{2})\\
(\nicefrac{1}{2},-\nicefrac{1}{2},-\nicefrac{1}{2},\nicefrac{1}{2}) &\xrightarrow{(12)(34)}& (-\nicefrac{1}{2},\nicefrac{1}{2},\nicefrac{1}{2},-\nicefrac{1}{2})\;,
\eea
\ese
they form a basis for the representation
\be
\Gamma^{\rm nu.sp}_{\rm reg,C_2} =  {\rm A}_{\chi} \oplus {\rm B}_{\chi}
\ee
in $\rm C_2(M)$. Likewise, the states  $(-\nicefrac{1}{2},\nicefrac{1}{2},-\nicefrac{1}{2},\nicefrac{1}{2})$ and $(\nicefrac{1}{2},-\nicefrac{1}{2},\nicefrac{1}{2},-\nicefrac{1}{2})$ form
a basis for 
\be
\Gamma^{\rm nu.sp}_{\rm reg,C^{\perp}_2} =  {\rm A}_{\theta} \oplus {\rm B}_{\theta}\;,
\ee
because
\bse
\bea
(-\nicefrac{1}{2},\nicefrac{1}{2},-\nicefrac{1}{2},\nicefrac{1}{2}) &\xrightarrow{(14)(23)(56)}& (\nicefrac{1}{2},-\nicefrac{1}{2},\nicefrac{1}{2},-\nicefrac{1}{2})\\
(\nicefrac{1}{2},-\nicefrac{1}{2},\nicefrac{1}{2},-\nicefrac{1}{2}) &\xrightarrow{(14)(23)(56)}& (-\nicefrac{1}{2},\nicefrac{1}{2},-\nicefrac{1}{2},\nicefrac{1}{2})\;.
\eea
\ese
Consequently, for all irreducible representations in $\rm C_2(M)$ and ${\rm C^{\perp}_2(M)}$, we find nuclear spin states of C$_2$F$_4$ that form a basis for these irreducible 
representations.

Here, we have limited the symmetry analysis of the nuclear spin states to subgroups of $\rm G^{psms}$ only. Properly identifying the observable nuclear spin isomers of a molecule, 
however, requires the use of the full permutation subgroup of the MS group.\autocite{Grohmann.2011,Grohmann.2012,Grohmann.2018b} Yet, in case $\rm G^{psms}$ is not cyclic, 
we can apply the method of (inverse) correlation to relate the symmetries of the nuclear spin states in the subgroups $\rm C_n(M)$ and ${\rm C^{\perp}_2(M)}$ with the irreducible 
representations of $\rm G^{psms}$.\autocite{Bunker.1998} By doing so, we are able to show that the nuclear spin functions transforming irreducible in $\rm C_n(M)$ and 
${\rm C^{\perp}_2(M)}$ can be associated with the nuclear spin isomers of the molecule. We return to this aspect at the end of next paragraph.

\paragraph{Localised states as superpositions of different nuclear spin isomer states}
The spin-statistics theorem allows us to identify each nuclear spin function with one nuclear spin isomer, because it only permits combinations of spatial states and nuclear spin 
states that fulfil \autoref{Eq:def-nsi}. Since these combinations are unique in $\rm G^{psms}$ (and its subgroups), each spatial state that transforms irreducible in $\rm G^{psms}$ 
represents a nuclear spin isomer of the molecule. 

Showing that localised states are superpositions of states representing different observable nuclear spin isomers is straightforward for molecules whose permutation subgroup of 
the MS group, $\rm G^{psms}$, is cyclic. A textbook example for this case is NF$_3$ with $\rm G^{psms}$ $=$ $\rm C_3(M)$, and $\mal I({\rm ^{14}N})=1$, 
$\mal I({\rm ^{19}F})=\nicefrac{1}{2}$; see \autoref{Fig:nf3} for three classical configurations. As \autoref{Eq:loc-NF3} shows, any of the localised states identified with these 
classical configurations is a superposition of rotational states of different symmetries in $\rm C_3(M)$. The spin states of the three fluorine nuclei span the
re\-pre\-sen\-ta\-tion\autocite{Bunker.1998,Grohmann.2010}\nolinebreak[4]
\be
\Gamma^{\rm nu.sp}
=
4{\rm A} \oplus  2{\rm E_1} \oplus 2{\rm E_2} \;.
\ee
Since any molecular state has to transform according to $\Gamma^{\rm mol}$ $=$ $\rm A$, NF$_3$ occurs in form of three nuclear spin isomers\autocite{Grohmann.2010}
\be
{\rm A}[{\rm A}] \quad {\rm E_1}[{\rm E_2}] \quad {\rm E_2}[{\rm E_1}]\;;
\ee
see \autoref{Eq:def-nsi}. Consequently, the three symmetry-adapted rotational functions forming the localised states in \autoref{Eq:loc-NF3} represent the three different 
nuclear spin isomers of NF$_3$.

An illustrative example for a molecule with permuted nuclei of zero spin is CO$_2$ with $\rm G^{psms}$ $=$ $\rm C_2(M)$, and $\mal I({\rm ^{12}C})=0$, $\mal I({\rm ^{16}O})=0$. 
For {CO$_2$}, only one totally symmetric nuclear spin state exist. Since any molecular state has to be totally symmetric too to fulfil the spin-statistics theorem, only totally 
symmetric rotational states are symmetry-allowed; see \autoref{Eq:def-nsi}. Hence, for {CO$_2$} localised states \textit{principally} cannot exist, because the existence of 
rotational states belonging to different $\Gamma^{\rm rot}$ in $\rm G^{psms}$ is a necessary condition for forming localised states.

If $\rm G^{psms}$ is not cyclic, but decomposes into cyclic subgroups, using the method of (inverse) correlation allows us to show that localised states are superpositions of different
nuclear spin isomer states. In case of C$_2$F$_4$, for example, the full permutation subgroup of the MS group is $\rm D_2(M)$ with irreducible representations ${\rm A}$, 
${\rm B_a}$, ${\rm B_b}$, and  ${\rm B_c}$.\autocite{Grohmann.2011} In $\rm D_2(M)$, the sixteen nuclear spin states of C$_2$F$_4$ span the representation
\be
\Gamma^{\rm nu.sp} 
=
7 {\rm A} \oplus 3 {\rm B_a} \oplus 3 {\rm B_b} \oplus 3 {\rm B_c} \;,
\ee
and \autoref{Eq:def-nsi} shows us that there are four observable nuclear spin isomers of C$_2$F$_4$,\autocite{Grohmann.2011}
\be
\label{Eq:nsi-C2F4}
{\rm A}[{\rm A}] \quad {\rm B_a}[{\rm B_a}] \quad {\rm B_b}[{\rm B_b}] \quad {\rm B_c}[{\rm B_c}]\;.
\ee
However, the symmetry-adapted states in \autoref{Eq:loc-wf-chi-1}, \autoref{Eq:loc-wf-chi-2}, \autoref{Eq:loc-wf-theta-1}, and \autoref{Eq:loc-wf-theta-2}, which form the localised 
states in $\chi$ and $\theta$, have been classified according to the subgroups $\rm C_2(M)$ and $\rm C^{\perp}_2(M)$. To identify how the nuclear spin isomers from 
\autoref{Eq:nsi-C2F4} are related to the spatial wave functions in \autoref{Eq:loc-wf-chi-1}, \autoref{Eq:loc-wf-chi-2}, \autoref{Eq:loc-wf-theta-1}, and \autoref{Eq:loc-wf-theta-2},
we have to use the method of correlation.\autocite{Bunker.1998} 

Correlating the irreducible representations of $\rm D_2(M)$ and $\rm C_2(M)$ gives\autocite{Grohmann.2011}
\bse
\bea
({\rm A},{\rm B_a}) &\rightarrow & {\rm A}_{\chi}\\
({\rm B_b},{\rm B_c}) &\rightarrow & {\rm B}_{\chi}\;.
\eea
\ese
Hence, rotational functions with $\rm A_{\chi}$-symmetry in $\rm C_2(M)$ represent either the nuclear spin isomer ${\rm A}[{\rm A}]$ or ${\rm B_a}[{\rm B_a}]$, while rotational 
functions with symmetry ${\rm B}_{\chi}$ in $\rm C_2(M)$ belong to the nuclear spin isomer ${\rm B_b}[{\rm B_b}]$ or ${\rm B_c}[{\rm B_c}]$. Analogously, we can correlate the 
irreducible representations of $\rm D_2(M)$ with the irreducible representations of $\rm C^{\perp}_2(M)$ to find\autocite{Grohmann.2011}
\bse
\label{Eq:D2-C2p-corr}
\bea
({\rm A},{\rm B_c}) &\rightarrow & {\rm A}_{\theta}\\
({\rm B_a},{\rm B_b}) &\rightarrow & {\rm B}_{\theta}\;.
\eea
\ese
\Autoref{Eq:D2-C2p-corr} shows that rotational functions with $\rm A_{\chi}$-symmetry in $\rm C^{\perp}_2(M)$ belong either to the nuclear spin isomer ${\rm A}[{\rm A}]$ or  
${\rm B_c}[{\rm B_c}]$, and rotational functions with symmetry ${\rm B}_{\theta}$ in $\rm C^{\perp}_2(M)$ represent the nuclear spin isomer ${\rm B_b}[{\rm B_b}]$ or 
${\rm B_c}[{\rm B_c}]$. Consequently, also in case $\rm G^{psms}$ decomposes into cyclic subgroups, localised states are superpositions of states belonging to different nuclear spin 
isomers.

\paragraph{No frozen nuclei for stable nuclear spin isomers}
Such superpositions, however, are not physical on the time-scales of molecular control experiments, simply because they cannot be created 
deliberately.\autocite{Grohmann.2018b} As the discussion from the preceding paragraphs shows, to create localised states we have to be able to superimpose rotational states of 
different symmetries $\Gamma^{\rm rot}$ in $\rm G^{psms}$ such that each symmetry-adapted state contributes with the same weight. Two strategies are possible to excite such 
states within the framework of the {frozen-nuclei approximation}: either by exciting rotational states of different symmetries directly with laser fields; or indirectly by 
superimposing nuclear spin states belonging to different $\Gamma^{\rm nu.sp}$ with magnetic fields. As we discuss in the following, both strategies fail.

The direct approach cannot work because of the permutational symmetry of molecules in electromagnetic fields. If \autoref{Eq:nuc-spin-hyp} is valid, exciting localised states is not 
possible, because any interaction that creates superpositions of states with different $\Gamma^{\rm rot}$ in $\rm G^{psms}$ must be able to break the permutational symmetry of 
the system. Electromagnetic fields only break the inversion symmetry, but permutations remain symmetry operations.\autocite{Watson.1975} Thus, the permutation subgroup of 
the MS group remains a symmetry group in the presence of the laser field, and the respective field-matter Hamiltonian can only couple rotational states belonging to the same 
irreducible representation in $\rm G^{psms}$.\autocite{Grohmann.2018b,Watson.1975} Mixing rotational states directly is therefore not possible if we assume that the rotational 
motions are adiabatically separated from other molecular motions. Consequently, the first strategy for creating localised states fails.

Although most studies on strong-field control assume \autoref{Eq:nuc-spin-hyp} to be valid, the frozen-nuclei approximation does not have to include the separation of nuclear spin 
states and spatial motions. It is possible to couple rotational motions with nuclear spins, if, for example, the dipolar interaction or the spin-rotation coupling is included in the 
description of the nuclei.\autocite{Chapovsky.1999,Grohmann.2010} Due to the symmetry properties of the molecular tensors describing these interactions, including them 
creates superposition states of different $\Gamma^{\rm nu.sp}$---and hence, of different rotational symmetries in $\rm G^{psms}$. This is the premise of the second strategy for 
creating localised states.

Usually, however, couplings of rotational motions and nuclear spins are very small, and nuclear spin conversion is an extremely slow process compared to the excitation of coherent 
molecular motions.\autocite{Chapovsky.1999,Grohmann.2010} Although nuclear spin converting effects can be amplified with magnetic fields,\autocite{Grohmann.2010} 
preparing superpositions of states belonging to different nuclear spin isomers is only possible on time-scales that are much longer than the time-scale on which molecular rotations 
typically take place. Hence, decoherence effects due to interactions with the environment\autocite{Schlosshauer.2008} would jeopardize any attempt to create localised 
states.\autocite{Chapovsky.1999,Grohmann.2010} Moreover, the superpositions created by the magnetic field do not fulfil the condition that every rotational state shall contribute
with equal weight.\autocite{Grohmann.2010,Grohmann.2018b} This follows from the fact that within a magnetic field, the MS group remains a true symmetry 
group.\autocite{Watson.1975} As a consequence, it is not possible to directly change the nuclear spin symmetry, but only indirectly by employing the dipolar interaction or the 
spin-rotation coupling. These couplings, however, do only occur between specific combinations of rotational and nuclear spin states, not between all of 
them.\autocite{Grohmann.2010} Thus, if at all, a magnetic field could only create a partial localisation. 

For these two reasons, the {frozen-nuclei-ap\-proxi\-ma\-tion} fails to describe nuclei that are confined in strong laser fields. To adequately simulate systems with 
hindered nuclear motion, we either have to use properly (anti-)symmetrised rotational states, or we need to go beyond the {frozen-nuclei approximation} and extend the present 
theories to explain why localised rotational states are reasonable descriptions for molecules with feasible permutations of identical nuclei. Both approaches, however, significantly 
deviate from the assumptions scientists make in the context of molecular control.

\subsection{A note on non-rigid molecules}
We insisted many times that the deliberations we make and the conclusions we draw in the present paper are limited to rigid molecules. Many interesting phenomena in molecular 
physics and chemistry, however, are related to molecules that cannot be described by the concepts used for rigid molecules, be it intramolecular motions and molecular rotors,
photochemistry, or charge-transfer phenomena. To theoretically describe these non-rigid molecules with large amplitude internal motions, many works rely on {frozen-nuclei 
approximations}, too.

Yet, treating non-rigid molecules theoretically is far more difficult than rigid molecules: neither a systematic analysis of their MS groups exists, nor the derivation of the zero-order 
Hamiltonian is standardized. For non-rigid molecules, different procedures for defining the molecule-fixed coordinate system have been 
developed.\autocite{Lin.1959,Bunker.1998,Ezra.1979,Ezra.1981,Ezra.1982,Papousek.1982,Hougen.2009,Woodman.1970,Soldan.1996} Depending on the method, the rotational 
and contortional degrees have different meanings, and the MS group might have to be modified if a specific method is used.\autocite{Bunker.2009,Hougen.2009,Soldan.1996} 
Both the diversity of methods and the sophisticated structure of their MS groups make general statements about the validity of the {frozen-nuclei approximation} for non-rigid 
molecules challenging.

Still, we can argue that at least full rotational localisation is not possible also for non-rigid molecules. Each MS group $\rm G^{\rm ms}$ of a molecule with large-amplitude internal 
motions can be expanded according to\autocite{Watson.1965}
\be
\label{Eq:MS-cosets}
{\rm G}^{\rm ms}
=
{\rm G}^{\rm eq}
\cup
\hat{\mal F}_2\,{\rm G}^{\rm eq}
\cup
...
\cup
\hat{\mal F}_{{\mak n}_{\rm V}}\,{\rm G}^{\rm eq}
=
\bigcup_{i = 1}^{{\mak n}_{\rm V}}\hat{\mal F}_i\,{\rm G}^{\rm eq}\;
\quad
\text{with}
\quad
\hat{\mal F}_1\equiv {\rm E}\;.
\ee
In \autoref{Eq:MS-cosets}, the group $\rm G^{\rm eq}$ is the MS group of the rigid energy minimum structure of the non-rigid molecule; it is isomorphic to the molecular point 
group of the electronic energy minimum structure.\autocite{Hougen.1962,Hougen.1963} The operators $\hat{\mal F}_i$ transform the reference version $1$ into version $i$, 
\textit{i.e.} they are permutations or permutation-inversions interconverting the ${\mak n}_{\rm V}$ versions of the molecule that are separated by superable energy 
barriers.\autocite{Watson.1965,Bunker.1998}

As the expansion \autoref{Eq:MS-cosets} shows, the MS group of the rigid energy minimum structure, $\rm G^{\rm eq}$, is a subgroup of the full MS 
group.\autocite{Watson.1965,Bunker.1998,Ezra.1982,Ezra.1979,Ezra.1981} Due to the isomorphism of $\rm G^{\rm eq}$ and the respective molecular point group, the arguments 
we present in \autoref{sec:sym-2step} and this section also apply to the group $\rm G^{\rm eq}$. This suggests that at least with respect to the operations of $\rm G^{\rm eq}$, 
localised rotational states are not possible for non-rigid molecules as well: symmetry-adapted rotational states, if classified according to the irreducible representations of the 
permutation subgroup of $\rm G^{\rm eq}$, represent different nuclear spin isomers of the non-rigid molecule, and superpositions of these states, such as localised states, are 
unphysical representations of the molecule.

The problem with this argument, however, is that for a non-rigid molecule, we cannot say with certainty what the meaning of the operations in \autoref{Eq:MS-cosets} is, because 
their interpretation strongly depends on the definition of the molecule-fixed coordinate system.\autocite{Lin.1959,Bunker.1998,Papousek.1982,Hougen.2009} Hence, solely from 
the structure of the MS group, we cannot conclude according to which irreducible representations the rotational wave functions transform. There are examples of which also the 
operations $\hat{\mal F}_i$ change the rotational coordinate, and not the contortional coordinate alone.\autocite{Bunker.1998,Soldan.1996} In a future publication, we will 
address this issue by discussing examples of contemporary research on molecular torsions.\autocite{Grohmann.2021} A related discussion on non-rigid molecules with observable
inversions was given recently.\autocite{Bouakline.2020}

\subsection{Freezing nuclei is not an approximation, but a theory reduction}
\label{subsec:theo-red}
When reading books about molecular physics and theoretical chemistry, it is almost impossible not to be confronted with the idea that the prominent approach of \textsc{Born} and 
\textsc{Oppenheimer} to calculating molecular eigenstates\autocite{Born.1924,Born.1927} is an approximation. Yet, in the literature on the philosophy of theoretical chemistry, it 
has long been argued that the adiabatic separation of nuclear and electronic motions means more than 
that.\autocite{Primas.1983,Primas.1985,Hagedorn.1988,Primas.1998,Weininger.1984,Sutcliffe.1992,Woolley.1976,Woolley.1985} Instead, the \textsc{Born-Oppenheimer} 
programme is better understood as a weak theory reduction from a quantum to a classical treatment of the nuclei. By reducing the theoretical description to classical mechanics 
and treating the nuclei as distinguishable objects, molecular structures arise as an emergent---classical---property of the molecule.\autocite{Primas.1983,Primas.1998}

The term approximation, these papers point out, is neither mathematically\autocite{Hagedorn.1980,Hagedorn.1988,Klein.1992,Kargol.1994} nor 
conceptionally\autocite{Primas.1983,Primas.1998} correct. Mathematically, the Born and Oppenheimer is an asymptotic expansion in terms of the ratio 
$\varsigma=(\nicefrac{m}{M})^{\nicefrac{1}{4}}$, where $m$ is the mass of an electron and $M$ is a mean nuclear mass of the molecular system. The point $\varsigma=0$ is singular 
and therefore leads to a discontinuous change in the description of the molecule. Instead of treating the nuclei as quantum objects, they are considered as classical particles in the 
limit of $\varsigma$ $\rightarrow$ 0, which ultimately allows us to speak of a molecular structure.\autocite{Primas.1983,Primas.1998} 

These arguments apply to the {frozen-nuclei approximation} alike.\autocite{Hagedorn.1988} Because the change from a localised rotational state to a classical 
configuration is discontinuous, it is not only incorrect to assume that it would smoothly converge into the classical configuration if only the external interaction is strong enough; 
see also \autoref{sec:sym-2step}, \autoref{subsec:loc-state-arg}. It also shows that the nuclei are considered qualitatively different. Since they are described as classical objects 
at this level of theory, neither we can motivate the existence of nuclear spin isomers, nor we can understand why localised states are physically wrong. Hence, if the 
{frozen-nuclei approximation} should be of any merit for describing experiments on molecular control, we need to make plausible why it is legitimate to describe the 
nuclei as (quasi-)classical objects.

\section{Conclusion: (anti-)symmetrisation is necessary---twice}
\label{sec:elec-antisym}
In this paper, we have shown that the {frozen-nuclei approximation} is an unphysical description of rigid molecules with identical nuclei. Because the quantum analogue of 
frozen nuclei, a localised rotational state, is a superposition of states belonging to different nuclear spin isomers of the molecule, it is either symmetry-forbidden, or it cannot be 
created on the time-scale of typical experiments on molecular control. The deeper reason for this inherent flaw is that the \textquote{frozen-nuclei approximation} is a theory 
reduction rather than an approximation, which treats the nuclei of the molecule as classical particles. Therefore, it cannot account for the indistinguishability of identical nuclei, the 
existence of nuclear spin isomers, and the impossibility of localised rotational states.

Consequently, for molecules with identical nuclei, the conventional approaches to molecular control need to be extended. An obvious example is the orbital tomography of 
symmetric molecules, like CO$_2$. The alleged reconstruction of molecular orbitals not only relies on the assumption that the electronic wave function decomposes into 
one-electron functions.\autocite{Scerri.2001,Schwarz.2006,Labarca.2010,Ogilvie.2011,Autschbach.2012,Villani.2017} It also assumes that it is justified to freeze the nuclei at their 
equilibrium configuration to eliminate the dependence of orbitals on the nuclear coordinates. As our argument shows, however, this assumption is wrong. Because nuclear wave 
functions have to obey the spin-statistics theorem, too, it is impossible to describe CO$_2$ by a localised rotational state. Instead, rotational states must be symmetry-adapted 
with respect to the operations of the permutation subgroup of the MS group, even in strong electromagnetic fields. Yet, symmetry-adapted states are necessarily delocalised in the 
rotational manifold, and hence, we have to take into account that minima corresponding to different potential-dressed versions are equally populated. As a consequence, we must 
not consider one particular classical configuration for the nuclei, but we only can, if at all, measure orbitals that are averaged over all potential-dressed versions of the molecule.

Another area of research for which the failure of the {frozen-nuclei approximation} is crucial is the contortional control of polyatomic 
molecules.\autocite{Grohmann.2018b,Grohmann.2018c} Models for describing internal motions in molecules, such as torsions or pseudo-rotations, often rely on the assumption 
that slow rotational motions can be considered as frozen over the course of the simulated contortional motion; see in particular Ref. \citenum{Grohmann.2018c} for an analysis and 
the references therein for examples. Many of the studied systems, however, exist in form of distinct potential-dressed versions, and applying the frozen-nuclei approximation is 
equivalent to choosing one of them and ignoring the consequences of the spin-statistics theorem. Hence, in particular in the context of directed motions, it is necessary to 
re-evaluate the premises of the models on contortional control to judge if they can account for the impossibility of localised rotational states.\autocite{Grohmann.2021b}

And yet, the fact that conventional models of molecular control fail in case they rely on the frozen-nuclei approximation is not the only conclusion we can draw from the 
impossibility of localised rotational states for molecules with stable nuclear spin isomers. Under the condition that the nuclei of a molecule can be treated as (quasi-)classical 
objects, our discussion shows, the frozen-nuclei approximation is appropriate. Hence, if it was possible to explain why such quasi-classical states of the nuclei exist, and why from all 
possible basis sets spatial position plays a special role,\autocite{Polkinghorne.2002} the localisation of molecules would appear as reasonable assumption. The environment-induced 
superselection of a preferred basis approach\autocite{Schlosshauer.2005,Schlosshauer.2008} might offer such explanation.

But no matter if spatial states are preferred or not, our conclusions show that results based on conventional models of molecular control employing the frozen-nuclei approximation 
have to be treated with care. For motions that can be described in terms of permutations of identical nuclei, models relying on localised states are ignorant about the quantum 
properties of nuclei. They cannot account for what might be decisive to properly understand the molecule under control: that nuclear wave functions have to obey the 
spin-statistics theorem, too.{\color{DFG-blau}\reflectbox{\ding{111}}}

\begin{center}
{\tikz[overlay,centered]{\draw [line width=1pt,-,line cap=round,color=darkgray] (-0.5\textwidth-0.125cm,0) -- (0.5\textwidth+0.125cm,0);}}
\end{center}

 \renewcommand{\thesection}{X}
\section{Appendix: Some useful characteristics of MS-groups}
\label{sec:append}
Some of the arguments we have presented in \autoref{sec:sym-2step} are not only valid for rigid molecules, but can be generalised to any type of MS group. Specifically, the 
typology of MS groups makes it easier to identify nuclear spin isomers in particular and allows for making conclusions about molecular observables in general.

\subsection{MS groups and point group operations}
\label{subsec:oper-ms}
To understand the typology introduced in \autoref{sec:sym-2step}, it is important to recall some basic rules on the relationship of the operations of the MS group and point group 
operations. 

For rigid molecules, to each permutation or permutation-inversion $\mal O^{\rm ms}$ in \autoref{Eq:op-MS} a different point-group operation $\mal O^{\rm pg}$  belongs. As 
shown by \citeauthor{Hougen.1962}, the complete set of operations constitutes the point-group of the rigid molecule;\autocite{Hougen.1962,Hougen.1963} see also the book of 
\citeauthor{Steinborn.1993}.\autocite{Steinborn.1993}

Only because of this isomorphism of the MS group and the point group of the rigid molecule, it is possible to map the operations of the MS group unambiguously on the operations
of the respective molecular point group. The rules for this mapping are as follows: Permutations $\mal P$ in the MS group correspond to proper rotation operations 
$\hat{\mal C}_{\rm n}$ in the molecular point group, while permutation-inversions $\mal P^*$ are equivalent to improper rotation operations $\hat{\mal S}_{\rm n}$. Additionally, 
proper and improper rotations of order $\rm n$ can be mapped only on permutations and permutation-inversions consisting of disjoint cycles with length $\rm n$, 
respectively.\autocite{Hougen.1962,Hougen.1963,Bunker.1998}

It is important to understand that $\hat{\mal C}_{\rm n}$ does \textit{not physically rotate} the molecule; it only rotates the vibronic \textit{coordinates} about the molecule-fixed 
$\bm e_z$-axis by the angle $\eta=\nicefrac{2\pi}{\rm n}$. Conversely, the equivalent rotations $\mal R^{\rm eq}$ change the rotational coordinates by rotating the molecule-fixed 
\textit{axes}. For rotations about the $\bm e_z$-axis, we can furthermore say: If $\mal R^{\rm eq}=\mal R^{\beta}_{z}$, see \autoref{Eq:R-eq-beta}, the corresponding point group 
operation rotates the vibronic {coordinates} by the angle $\eta=2\pi-\beta$ about the molecule-fixed $\bm e_z$-axis.\autocite{Bunker.2005}

For improper rotation operations, the following rules hold: $\hat{\mal S}_1 \equiv \hat{\sigma}_{\rm mol}$ and $\hat{\mal S_2} \equiv \hat{i}$. Here, $\hat{\sigma}_{\rm mol}$ is a 
mirror pane being perpendicular to the highest axis of rotation. For any improper rotation of higher order, $\hat{\mal S}_{\rm n}=\hat{\mal C}_{\rm n}\cdotp\sigma_{\rm mol}$, 
with the rules for the operations $\hat{\mal C}_{\rm n}$ being specified in the preceding paragraph. We stress that the inversion $\rm E^*$ does \textit{not} correspond to the 
inversion operation in the point group, but to the reflection at the molecular plane $\hat{\sigma}_{\rm mol}$.\autocite{Bunker.2005} 

Further, these rules are only valid for rigid molecules. For non-rigid molecules, a permutation $\mal P$ or a permutation-inversion $\mal P^*$ might only change the contortional 
coordinates, leaving the rotational and vibronic coordinates of the molecule unaffected. Moreover, the meaning of the operations of the MS group in terms of molecular 
coordinates, in general, depends on which method is used to set up the {zero-order} Hamiltonian. Unlike for rigid molecules, the molecule-fixed coordinate system is not defined 
uniquely; different definitions of this coordinate system lead to different geometrical interpretations of the operations.\autocite{Bunker.1998,Hougen.2009}

\subsection{A typology of MS groups}
\label{subsec:types-ms}
As pointed out in \autoref{subsec:perm-sym-sym-top} of \autoref{sec:sym-2step}, any MS group falls into one of the following three categories: \textit{type-I}, 
\textit{type-II\textsuperscript{a}}, or \textit{type-II\textsuperscript{b}}. This typology is based on how the permutation subgroup $\rm G^{psms}$ relates to the full MS group of the
molecule. 

For MS groups of \textit{type-I}, this relation is simple: $\rm G^{psms}$ is an improper subgroup of $\rm G^{ms}_ {\rm I}$, because these MS groups do not contain any 
permutation-inversions. For MS groups of \textit{type-II\tssc{a}}, we stated that they can be written as a direct product of the permutation subgroup and the inversion group, while 
for MS groups of \textit{type-II\tssc{b}} this is not possible. As we point out in the following, their structure is a direct consequence of the inversion $\rm E^*$ being a feasible 
operation or not.

For MS groups of \textit{type-II\tssc{a}}, the inversion $\rm E^*$ is feasible. If $\rm E^*$ is feasible, the MS group must contain for each permutation $\mal P$ the corresponding 
permutation-inversion $\mal P^*$. For the following reason: Any permutation-inversion $\mal P^*$ can be written as 
\be
{\mal P}^* = {\mal P} {\rm E^*} = {\rm E}^*{\mal P}\;,
\ee
where $\mal P$ is a pure permutation of identical nuclei. If $\rm E^*$ is feasible, the element
\be
{\rm E^*}{\mal P}^*  =  {\rm E^*}{\rm E}^*{\mal P}={\mal P}
\ee
must also be in the MS group, otherwise the group axioms would not be fulfilled.\autocite{Wigner.1959} Hence, since $\mal P^*$ is an arbitrary but feasible permutation-inversion, 
its corresponding permutation $\mal P$ must be feasible, too. Since furthermore $\{\rm E,\rm E^*\}$ and $\rm G^{psms}$ both form normal subgroups of $\rm G^{\rm ms}$, and 
both have only the identity $\rm E$ in common, the complete MS group can be written as \autoref{Eq:MS-typ-IIa}.\autocite{Mirman.2007}

In MS groups of  \textit{type-II\tssc{b}}, the inversion $\rm E^*$ is not feasible. Here, the MS group contains for none of the feasible permutations $\mal P$ the corresponding 
permutation-inversion $\mal P^*$. Per definition, the only way of obtaining $\mal P^*$ from $\mal P$ is by combining $\mal P$  and $\rm E^*$.\autocite{Bunker.1998} Yet, if 
$\rm E^*$ is not feasible, it is not possible to obtain $\mal P^*$ by combining $\mal P$ with any element of the MS group. Thus, if $\mal P^*$ is feasible, for MS groups of 
\textit{type-II\tssc{b}}, $\mal P$ is not, and \textit{vice versa}. Nevertheless, the permutation subgroup $\rm G^{psms}$ still forms an invariant subgroup of the MS group, but it 
can no longer be written as a direct product, see \autoref{Eq:MS-typ-IIb}.

Since the arguments are independent of the geometrical realization of the operations of the MS group, the results apply to any MS group, irrespectively of the molecule being rigid 
or non-rigid.

\subsection{On the irreducible representations of MS groups}
\label{subsec:irreps-ms-types}

The structure of MS groups has direct consequences for the labelling of their irreducible representations. If $\Gamma_{\alpha}$ specifies an irreducible representation of the 
permutation subgroup, for the irreducible representations of the MS group holds the following:

\setlength\mathindent{8.25em}
\usetagform{brackets4}

\bit
[leftmargin=4em,
topsep=0\baselineskip,
itemsep=0.5\baselineskip,
format={\it\color{DFG-blau}},
labelwidth=3.75em,
align=left
]
\item[{\color{DFG-blau} Type-I}] 
The irreducible representations of the MS group can be chosen to be identical to $\Gamma_{\alpha}$.

\item[{\color{DFG-blau} Type-II$^a$}] 
The irreducible representations of $\rm G^{\rm ms}$ can be labelled according to $\Gamma_{\alpha,\pm}$. Here, for the characters of 
$\Gamma_{\alpha,\pm}$,\autocite{Bunker.1998}
\be
\zeta^{\Gamma_{\alpha,\pm}}[{\mal P^*}]
=
\pm\zeta^{\Gamma_{\alpha,\pm}}[{\mal P}]\;,
\ee
and no extra degeneracies are introduced by including the inversion operation $\rm E^*$. Moreover, the irreducible representations of the MS group correlate with the 
irreducible representations of its permutation subgroup according to
\be
\Gamma_{\alpha,\pm}
\xrightarrow{{\rm G^{\rm ms}}\, \rightarrow\,{\rm G^{\rm psms}}} \Gamma_{\alpha} \;.
\ee

\item[\color{DFG-blau} Type-II$^b$] 
There is no general correlation scheme between $\Gamma_{\alpha}$ and the irreducible representations of the MS group. In particular, it is possible that two (or more) irreducible 
representations of $\rm G^{\rm psms}$ correlate with only one irreducible representation in the MS group. Therefore, including permutation-inversions might introduce additional 
degeneracies.
\eit
\setlength\mathindent{4.25em}
\usetagform{brackets2}

For identifying nuclear spin isomers, this is in particular important. Often, the states of the molecular system can be classified in the MS group, but only using the permutation 
subgroup allows for uniquely defining the nuclear spin isomers of molecules. While for molecules belonging to MS groups of \textit{type-I} or \textit{type-II$^a$} it is still possible 
to uniquely define the nuclear spin isomers of the molecule in the MS group, for groups of \textit{type-II$^b$} this is no longer possible. To identify the nuclear spin isomers of a 
molecule belonging to MS groups of \textit{type-II$^b$}, we then have to use the method of inverse correlation,\autocite{Bunker.1998,Grohmann.2011,Grohmann.2012}; see Ref. 
\citenum{Grohmann.2011} and \citenum{Grohmann.2012} for examples.

\begin{center}
{\tikz[overlay,centered]{\draw [line width=1pt,-,line cap=round,color=darkgray] (-0.5\textwidth-0.125cm,0) -- (0.5\textwidth+0.125cm,0);}}
\end{center}
\vspace*{-0.75\baselineskip}

\renewcommand\refname{\bfseries\Large References\vspace*{-0.15\baselineskip}}
\renewcommand{\leftmark}{\small References}
\singlespacing
\renewcommand{\bibfont}{\normalfont\footnotesize}
\addcontentsline{toc}{section}{{}{\bfseries References}}
\printbibliography

\end{document}